\documentclass[iop]{emulateapj}

\newcommand*{\teff}{$T_{\rm eff}$}
\newcommand*{\logg}{$\log~g$}
\newcommand*{\feh}{[Fe/H]}
\newcommand*{\afe}{[$\alpha$/Fe]}
\newcommand*{\kms}{km s$^{-1}$}
\newcommand*{\zmax}{$Z_{\rm max}$}
\newcommand*{\rapo}{$r_{\rm a}$}
\newcommand*{\rperi}{$r_{\rm p}$}

\newcommand*{\vrd}{$V_{\rm R}$}
\newcommand*{\vph}{$V_{\rm \phi}$}
\newcommand*{\vz}{$V_{\rm Z}$}

\newcommand*{\z}{$|Z|$}
\newcommand*{\kmskpc}{km s$^{-1}$~kpc$^{-1}$}
\newcommand*{\kmsdex}{km s$^{-1}$~dex$^{-1}$}

\newcommand*{\gaia}{$Gaia$}

\def\jnlref#1{{\rm#1}}
\def\apj{\jnlref{ApJ}}
\def\apjs{\jnlref{ApJS}}
\def\aap{\jnlref{A\&A}}
\def\mnras{\jnlref{MNRAS}}
\def\natur{\jnlref{Nature}}

\usepackage{color}
\usepackage{epsfig}
\usepackage{graphicx}
\usepackage{natbib}
\usepackage{hyperref}
\usepackage{amsmath}
\usepackage{epsfig}

\hypersetup{
    bookmarks=true,         
    unicode=false,          
    pdftoolbar=true,        
    pdfmenubar=true,        
    pdffitwindow=true,     
    pdfstartview={FitH},    
    pdftitle={},    
    pdfauthor={Lee},     
    pdfsubject={Astronomy},   
    pdfcreator={dvipdf},   
    pdfproducer={dvipdf}, 
    pdfkeywords={metal-poor stars},
    pdfnewwindow=true,      
    colorlinks=true,       
    linkcolor=red,          
    citecolor=blue,        
    filecolor=magenta,      
    urlcolor=cyan,           
    breaklinks=true,
    linktocpage
}

\shorttitle{Formation and Evolution History of the Galactic Disk System}
\slugcomment{Draft version, May 7, 2020}
\shortauthors{Han et al.}

\begin{document}
\title{Insights into the Formation and Evolution History of the Galactic Disk System}

\author{Doo Ri Han\altaffilmark{1}, Young Sun Lee\altaffilmark{2,4}, Young Kwang Kim\altaffilmark{2}, and Timothy C. Beers\altaffilmark{3}}
\altaffiltext{1}{Department of Astronomy, Space Science, and Geology, Chungnam National University,
                 Daejeon 34134, South Korea}
\altaffiltext{2}{Department of Astronomy and Space Science, Chungnam National University, Daejeon 34134, South Korea}
\altaffiltext{3}{Department of Physics and JINA Center for the Evolution of the Elements, University
                 of Notre Dame, Notre Dame, IN 46556, USA}
\altaffiltext{4}{Corresponding author: youngsun@cnu.ac.kr}

\begin{abstract}

We present a kinematic analysis of a sample of 23,908 G- and K-type dwarfs 
in the Galactic disk. Based on the $\alpha$-abundance ratio,
[$\alpha$/Fe], we separated our sample into low-$\alpha$ thin-disk and
high-$\alpha$ thick-disk stars. We find a \vph\ gradient of --28.2
\kmsdex\ over \feh\ for the thin disk, and an almost flat trend of the
velocity dispersions of \vrd, \vph, and \vz\ components with [Fe/H]. The
metal-poor (MP; [Fe/H] $<$ --0.3) thin-disk stars with low-\vph\
velocities have high eccentricities ($e$) and small perigalacticon
distances (\rperi), while the high-\vph\ MP thin-disk stars possess low
$e$ and large \rperi. Interestingly, half of the super metal-rich ([Fe/H] $>$ $+$0.1) 
stars in the thin disk exhibit low-$e$, solar-like
orbits. Accounting for the inhomogeneous metallicity distribution of the
thin-disk stars with various kinematics requires radial migration by
churning -- it apparently strongly influences the current structure of
the thin disk; we cannot rule out the importance of blurring for the
high-$e$ stars. We derive a rotation velocity gradient of $+$36.9
\kmsdex\ for the thick disk, and decreasing trends of velocity
dispersions with increasing \feh. The thick-disk population also has a
broad distribution of eccentricity, and the number of high-$e$ stars
increases with decreasing \feh. These kinematic behaviors could be the
result of a violent mechanism, such as a gas-rich merger or the presence
of giant turbulent clumps, early in the history of its formation.
Dynamical heating by minor mergers and radial migration may also play
roles in forming the current thick-disk structure.

\end{abstract}



\section{Introduction}

It has been almost four decades since the existence of the thick disk of
the Milky Way (MW) was established by fitting double-exponential
functions to the vertical density profile of stars in the Galactic disk
(\citealt{Yoshii82,Gilmore83}). Tremendous progress in understanding the
nature of the Galactic disk system has been made, in particular
recently, thanks to the advent of large photometric and spectroscopic
surveys such as the Sloan Digital Sky Survey (SDSS; \citealt{York00}),
the RAdial Velocity Experiment (RAVE; \citealt{Steinmetz06}), the Sloan
Extension for Galactic Understanding and Exploration (SEGUE;
\citealt{Yanny09}), the $Gaia$-ESO survey \citep{Gilmore12}, the Large
sky Area Multi-Object Fiber Spectroscopic Telescope (LAMOST;
\citealt{Luo15}), the Apache Point Observatory Galactic Evolution
Experiment (APOGEE; \citealt{Majewski17}), and the GALactic Archaeology
with HERMES (GALAH; \citealt{DeSilva15}). 

The properties of the thick disk of the MW differ in many aspects from
those of the thin disk. Briefly summarizing: (1) Spatially, the scale
heights of the thin- and thick-disks are about 300 pc and 900 pc,
respectively \citep{Juric08} (see also \citealt{Bland16} for a nice
review on the scaleheight of each disk); (2) Kinematically, the orbital
rotation velocity of the thick disk lags by about 30 \kms\ behind that
of the thin disk (\citealt{Lee11b,Jing16}), and the velocity dispersions
of the thick disk are larger than those of the thin disk
(\citealt{Blanco14,Guiglion15,Wojno16}); (3) Chemically, the
metallicity distribution function (MDF) of the thick-disk population
peaks at between [Fe/H] = --0.5 and --0.6, while that of the thin disk
is at about [Fe/H] = --0.2 \citep{Wyse95,Soubiran03,Kordopatis11,
Lee11b}. The $\alpha$-element abundance ratio with respect to Fe is
higher by 0.2 to 0.3 dex for the thick disk than for the thin disk
(\citealt{Lee11b,Adibekyan13,Boeche13,Anders14, Hayden15,Yan19}).

Note that, even though the disk system can be separated by the
properties described above, an alternative interpretation has been
advanced, that the thick disk is smoothly connected to the thin disk,
rather than existing as an independent component (e.g.,
\citealt{Bovy12a,Bovy12b}). It has also been reported that the
low-metallicity tail of the thick-disk MDF is an independent component,
known as the metal-weak thick disk (MWTD). This component has lower
metallicity (by at least a factor of two) and lower rotation velocity
(by about 30 \kms) than the canonical thick disk
\citep{Carollo19}. The independent nature of the MWTD has been further
confirmed with large photometric survey data combined with accurate
proper motions \citep{An20}. On the other hand,
\citet{Adibekyan11} reported, from an analysis of 1112 F, G, and K-type
dwarfs, that in addition to the canonical thick-disk stars with enhanced
\afe, metal-rich thick-disk stars ([Fe/H] $>$ --0.2) with enhanced
\afe\ may exist as a separate component. 

These observed differences between the two disk components strongly
suggest that they have undergone different formation and evolutionary
histories, as has been noted by many authors. \citet{Abadi03}, for
example, used numerical simulations to propose that the MW has undergone
a major merger at high redshift ($z$), and a large fraction of stars
have been accreted to form the thick disk from disrupted satellites
during the merging process. According to this scenario, more than 60\%
of the thick-disk stars are comprised of debris of the tidally-disrupted
satellite. \citet{Villalobos08} reported from their MW-like simulations
that a {\it minor} merger, for which the mass of the satellite accounts
for only 10 -- 20\% of that of the host galaxy, led to the thick-disk
formation. In this mechanism, the satellite galaxy serves to heat and
tilt the disk of the host galaxy without destroying it, resulting in
thickening a pre-existing thin disk. 

Recently, \citet{Helmi18} demonstrated that a 1:4 mass ratio merger,
referred to as \gaia-Enceladus, could have produced the thick disk by
dynamical heating of the pre-existing thin disk. The \gaia-Sausage may
have affected the formation of the thick disk in a similar way
\citep{Belokurov18}. Studies by \citet{Brook04,Brook07,Brook12} showed
that the thick disk could also be formed from gas-rich building blocks,
which were accreted during chaotic phase of hierarchical clustering at
high redshift, while \citet{Bournaud09} suggested that the thick disk
formed from giant turbulent clumps, which undergo gravitational
contraction and scatter stars at high $z$. All of these scenarios are
associated with violent events for the formation of the thick disk.

In contrast to these violent scenarios, other studies have claimed that
the formation of the thick disk could instead be the result of the
cumulative effects of secular process associated with the radial
migration of stars (\citealt{Sellwood02,Schonrich09,Loebman11,
Roskar12}). According to these authors, stars in the disk can
effectively migrate radially inward or outward through the so-called
$churning$ and $blurring$ processes. The $churning$ process arises from
the exchange of angular momentum stars with transient spiral structures;
riding these spiral patterns, stars can migrate significant distances 
from their birth place within the disk. As stars move from the inner to 
the outer regions of the disk, they experience a weaker gravitational 
potential due to the decreasing mass density with increasing distance from 
the Galactic center, which results in larger vertically excursions from the 
plane. On the other hand, the $blurring$ process is caused by the change of 
the epicycle amplitude of a star around its guiding center, without changing
its angular momentum, through perturbations by giant molecular clouds.
Some studies, however, have argued that the effect of radial migration
may not be sufficient to significantly thicken the disk (e.g.,
\citealt{Minchev12,Kawata17}).

As each formation model above predicts different (though perhaps not
unique) characteristics of the resulting thick disk, there have been
various efforts carried out to unravel the complex formation
mechanism(s) through comparisons with the theoretical predictions. For
example, \citet{Sales09} noted distinct distributions of the stellar
orbital eccentricities associated with various scenarios, and suggested
that the use of the distribution of observed orbital eccentricities for
the thick disk could help distinguish one from another. Although
numerous studies have employed a similar approach, using different
stellar samples to explore the various formation mechanisms, they have
thus far failed to pin down a unique model (\citealt{Wilson11,
Dierickx10,Casetti11,Lee11b,Li17,Yan19}).

In order to provide more stringent constraints on the suggested
formation scenarios for the thick disk, one first needs to separate
stars of the disk system into the thin- and thick-disk components, and
investigate the detailed spatial, kinematic, and chemical properties of
the stars belonging to each component. There have been several attempts
to divide the disk stars into components by relying on their observed
space motions (e.g., \citealt{Bensby03,Bensby14,Jing16}). If a star is
presently located in the same position as where it was born, and has the
same kinematic properties as its natal molecular cloud, kinematic
information could well be a powerful tool for this exercise. However, 
as it has become increasingly clear through investigations carried out with
ever larger stellar samples, that is not likely to be the case. The
birth places and kinematic properties of stars more likely change over
time due to either major or minor mergers or secular evolution via
perturbations by transient spiral patterns and giant molecular clouds.

By way of contrast, the chemistry of solar (and later-type) dwarf stars
is essentially invariant during their main-sequence lifetimes. Thus, it
provides a more stable and reliable indicator with which to classify
disk-system stars into various components, provided that such components
possess distinct chemical signatures. Among the various
elemental-abundance ratios, the alpha-element-to-iron ratios (\afe) have
proven useful for this exercise, as the \afe\ ratios are larger by 0.2 -- 0.3 dex 
for stars in the thick disk than for those in the thin disk
(see \citealt{Bensby03,Reddy06,Fuhrmann98,Fuhrmann08}). A
number of studies (\citealt{Lee11b,Adibekyan13,Blanco14,Wojno16,
Hayden15,Hayden18,Yan19}) have been performed to investigate the
kinematic properties of the stellar populations identified in the \feh\ 
versus \afe\ plane, using the large amount of data from various
spectroscopic surveys. Even though these studies were successful, to
some degree, in reproducing hotter kinematic properties of the thick
disk as well as the kinematic features of the thin disk, the formation
mechanism of the thick disk has remained resistant to consensus. 

This work is another effort to understand the formation mechanism of the
thick disk. In this study, we utilize SEGUE G- and K-dwarfs in the
Galactic disk, in order to explore their properties and compare with
various thick-disk formation models. We make use of \afe\ and \feh\ to
chemically separate disk-system stars into the thin- and thick-disk
populations, then examine the \vrd, \vph, \vz\ velocity components,
their gradients with metallicity, their velocity dispersions, and the
dynamical properties of each disk population. Based on our findings, we
suggest likely formation and evolution histories of the Galactic disk
system.

This paper is organized as follows. In Section \ref{sec:data_method}, we
describe the sample of G and K dwarfs, calculations of the velocity
components and orbital parameters, and the method for the chemical
separation of each disk population. We also discuss potential
selection biases in our sample. The results of the analyses of the
kinematic and dynamical properties of each stellar population are
presented in Section \ref{sec:result_vel} and \ref{sec:result_orbit},
respectively. We consider the plausible formation and evolution history
of each disk in Section \ref{sec:discussion}. Finally, we summarize our
work in Section
\ref{sec:summary}.

\section{SEGUE G and K Dwarfs}\label{sec:data_method}
\subsection{Sample Selection}

We first gathered stars that were specifically targeted as G and K
dwarfs in SEGUE, in the dereddened color and magnitude ranges of 0.45 $<
(g - r)_0 <$ 0.80 and 13.5 $< r_0 <$ 20.5, respectively. From application
of the SEGUE Stellar Parameter Pipeline (SSPP; \citealt{Allende08,
Lee08a,Lee08b,Lee11a,Smolinski11}) on the medium-resolution ($R \sim$ 2000)
spectra of those dwarfs, we determined the stellar atmospheric
parameters (\teff, \logg, and \feh) for each star. Typical errors of
these estimated parameters from the SSPP are $\sim$ 180 K for \teff,
$\sim$ 0.24 dex for \logg, and $\sim$ 0.23 dex for \feh. We also derived
\afe\ using the method of \citet{Lee11a}; the uncertainty of the
estimated \afe\ is typically better than 0.1 dex for signal-to-noise
ratio (S/N) larger than 20. For reliable stellar parameters and \afe, we
applied this S/N cut to the entire sample.

\begin{figure}[t]
\centering
\plotone{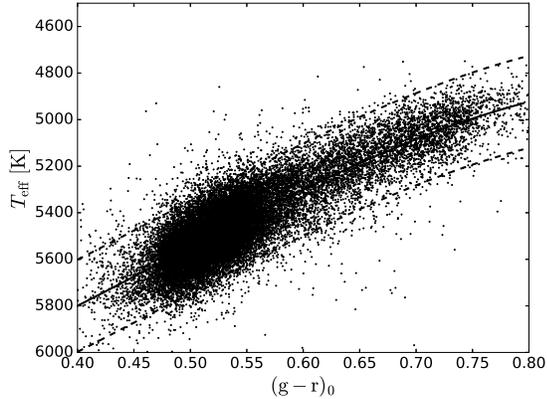}
\caption{Distribution of candidate main-sequence dwarfs in the
color-temperature plane. The black-solid curve is obtained from a
least-square fit to the data; the dashed lines indicate $\pm$2$\sigma$
deviations from the fitted line. Stars located outside the dashed lines
are removed from the sample in the remaining analysis.}
\label{fig:1}
\medskip
\end{figure}

Additionally, only stars with 4500 $<$ \teff\ $<$ 6000 K and \logg\ $>$
3.8 were included in our sample, in order to secure the most reliable
identification of G and K dwarfs. As a final check, adopting a
color-temperature relation for main-sequence stars, we removed stars
more than 2$\sigma$ above or below the relation at a given $(g - r)_0$,
as illustrated in Figure \ref{fig:1}. The black-solid line in the figure
comes from a least-square fit to the data, while the dashed lines are
the $\pm$2$\sigma$ limits from this line. Even though 
this 2$\sigma$ limit is somewhat arbitrary, its purpose is to remove 
stars with erroneously determined \teff\ (usually due to some defect 
in their spectra).

\begin{figure}[t]
\centering
\plotone{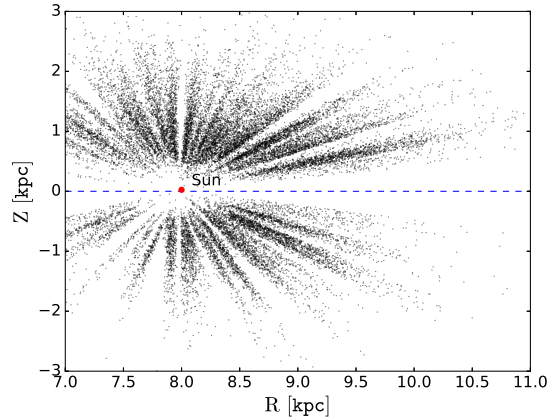}
\caption{Spatial distribution of the final program sample in the $R$ versus $Z$ plane, 
which includes 23,908 SDSS/SEGUE G and K dwarfs. Note that the cuts applied as
described in the text have already been applied here. The Sun (red dot)
is located at ($R$,$Z$) = (8.0,0.025) kpc; the blue dashed line at $Z$ =
0 kpc is the Galactic mid-plane.}
\label{fig:2}
\medskip
\end{figure}

\subsection{Calculations of Space Velocity Components and Orbital Parameters}

In order to calculate the most precise velocity components for our
program stars, we cross-matched our program stars with \gaia\ Data
Release 2 (DR2) \citep{GaiaColl18} in order to retrieve their proper
motions and parallaxes. We selected stars with listed uncertainties in
proper motion less than 1.0 mas yr$^{-1}$, and relative parallax errors
smaller than 25\% (=$\sigma_{\pi}/\pi$), where $\pi$ is the parallax and
$\sigma_{\pi}$ is the uncertainty. We also adjusted the parallaxes by
the zero-point offset --0.029 mas (\citealt{Lindegren18}). We computed
the distance by taking the inverse of the parallax for each star, and
calculated the radial distance from the Galactic center projected onto
the Galactic plane ($R$), as well as the vertical distance from the
Galactic plane ($Z$). We assumed that the Sun is located at $R$ = 8 kpc
and $Z$ = 0.025 kpc \citep{Bland16}. 

For a sanity check on the adoption of inverse-parallax distances
from the \gaia\ catalog, we compared our derived distances with those
estimated by a probabilistic inference approach \citep{BailerJones18}.
We found from the sample of stars in common that the mean difference is
--0.034 kpc, with a standard deviation of 0.072 kpc. Although the mean
offset becomes larger for more distant objects, the relative difference
is less than 20\% at 3 kpc, which is the typical uncertainty in
estimating distances. Because our dwarf stars are predominantly located
at a distance less than 3 kpc from the Sun, this offset suggests that
the distance derived from taking the inverse of the \gaia\ parallaxes is
suitable for our study. We note, however, that our derived 
distances may suffer from the Lutz-Kelker bias(\citealt{Lutz73}), which causes 
an underestimate of our derived distance, especially for more distant
objects, as they have larger relative errors ($\sigma_{\pi}/\pi$). 

Furthermore, by making use of the distances and proper motions, and adopting
radial velocities from SEGUE spectra, we derived the space velocity
components and orbital parameters, using the ${\tt MWPotential2014}$
from ${\tt galpy\ Python}$
package\footnote{$http://jobovy.github.io/galpy$.} \citep{Bovy15}. This
potential consists of a power-law density bulge and a Miyamoto-Nagai
disk with mass 6.8 $\times\ 10^{10}\ M_{\odot}$, and a
Navarro-Frenk-White dark-matter halo profile.

We first derived $U,~V,~W$ velocity components of motion for each star,
taking the peculiar motion of the Sun to be $(U,V,W)_\odot$ =
(--10.1,4.0,6.7) $\pm$ (0.5,0.8,0.2)
\kms\ \citep{Hogg05}, and the circular velocity of the local standard of rest (LSR)
to be 220 \kms\ \citep{Kerr86}. The positive direction of the $U$ component is
taken to be radially outward from the Galactic center. Solar orbital
rotation is taken to be positive in the $V$ direction, and $W$ is
positive in the direction of the North Galactic Pole. Additionally, the
velocity components (\vrd, \vph, \vz) in a cylindrical coordinate system
were computed for each star. Among the orbital parameters, we derived
the minimum distance (\rperi) and maximum distance (\rapo) from the
Galactic center, and the maximum distance (\zmax) from the Galactic
plane reached during the orbit of a given star. The eccentricity ($e$)
is obtained from $e =$ (\rapo$-$\rperi)/(\rapo$+$\rperi).

\begin{figure*}
\centering
\plotone{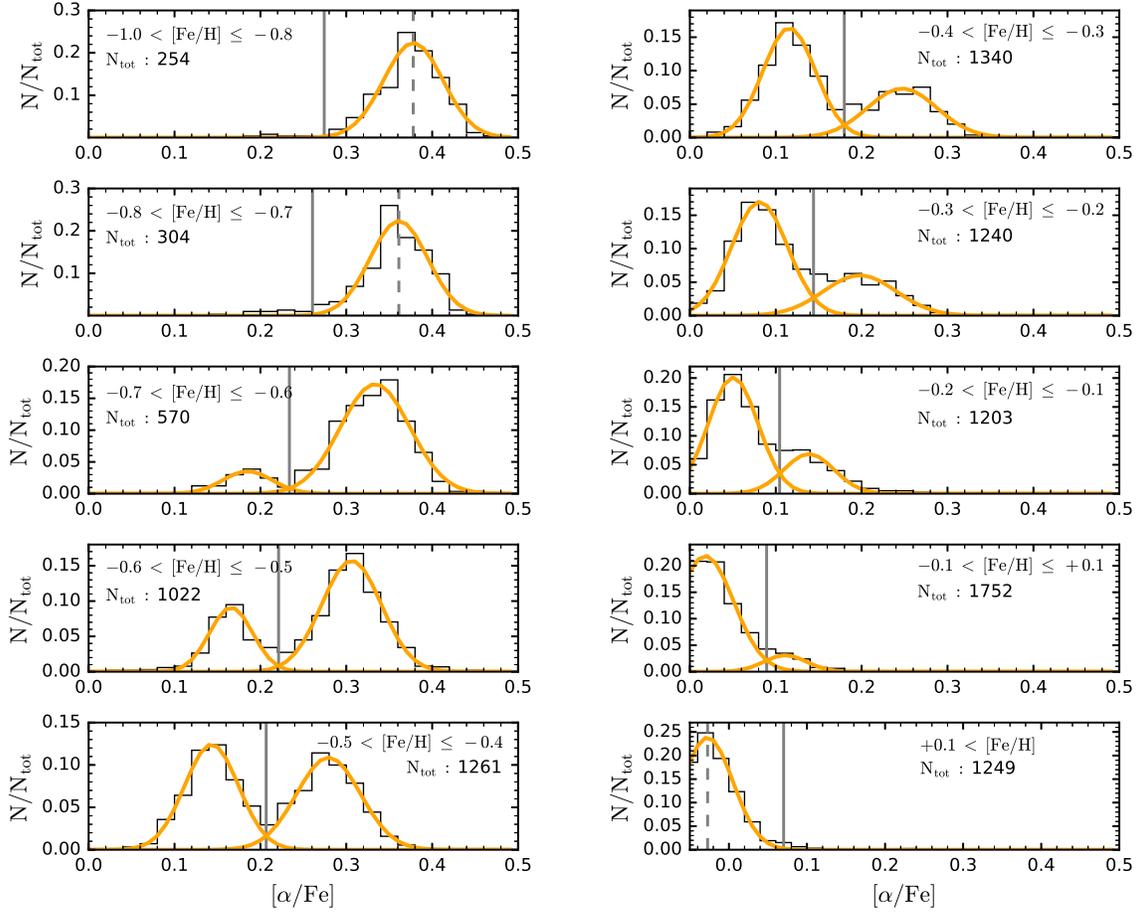}
\caption{Histograms of \afe\ for different \feh\ cuts. The black lines
are the original data. The orange curves are derived from one or two
Gaussian fits to each histogram. If two components exist, we use an
iterative procedure to obtain division points (gray solid lines), as 
described in the text, between the two components. Gaussian models
are then fit to the individual components. Note that in the 
range of --0.4 $<$ [Fe/H] $\leq$ --0.1, we first fit a Gaussian 
to the thin-disk component to find the best-fit parameters; after 
fixing these, we derived the best-fit Gaussian for 
the thick-disk population. For this procedure, we only
consider stars with spectra having S/N $>$ 50, in order to ensure a
robust split. $N_{\rm tot}$, listed in the legends, is the total number
of stars in each panel.}
\label{fig:3}
\end{figure*}

\begin{figure*}
\centering
\plottwo{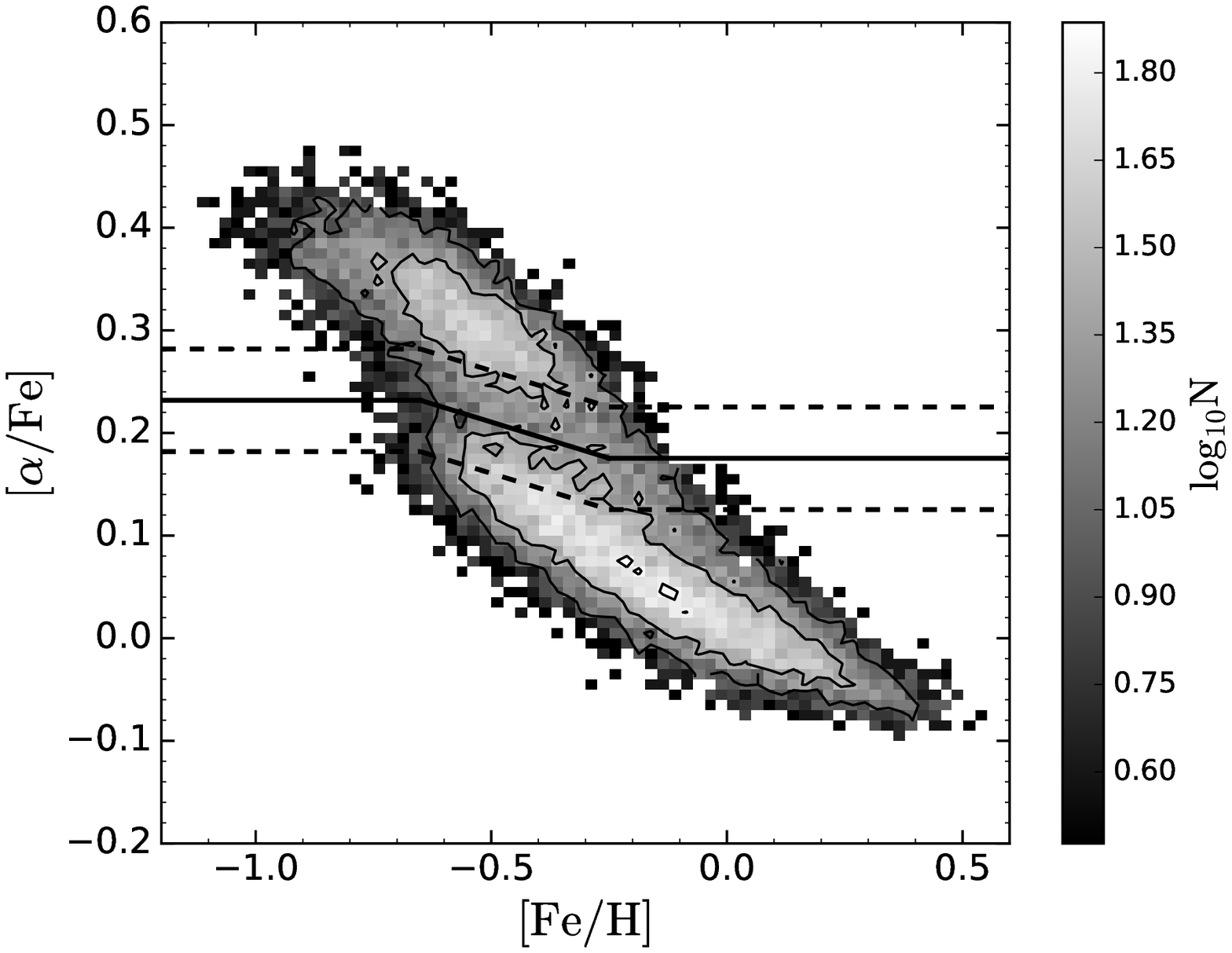}{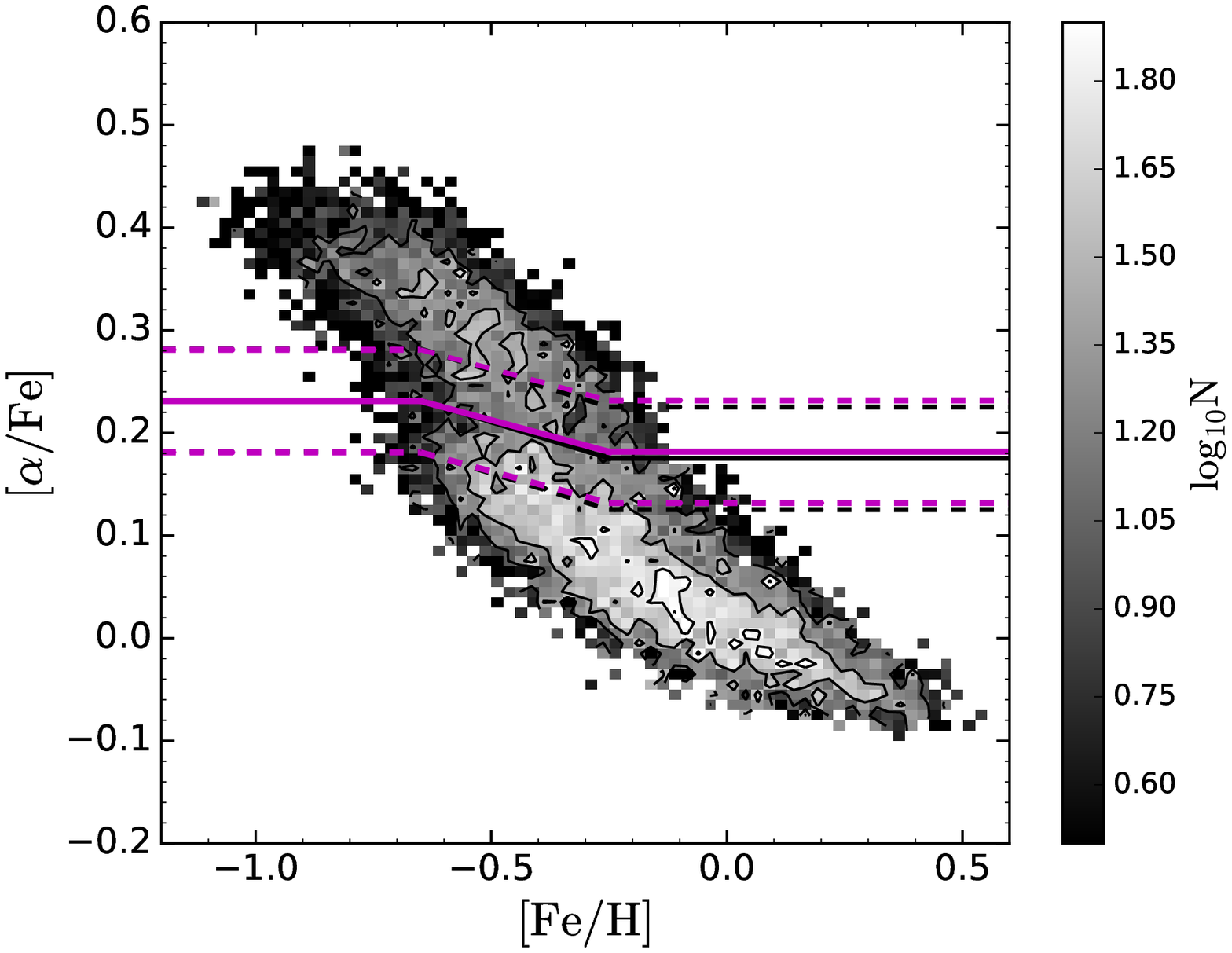}
\caption{Left panel: Logarithmic gray-scale plot of the number density in the
\feh\ versus \afe\ plane, overplotted with equal density contours. Each bin 
has a size of 0.025 by 0.01 in \feh\ and \afe, respectively, and
contains at least two stars. The black-solid line is a reference point
to aid in the separation of stars into the thin- and thick-disk
components. The dashed lines are located at $\pm$0.05 dex in \afe\ from
the solid line (see text for further discussion). Right panel: Same as
in the left panel, but for the selection bias-corrected distribution as described
in the text. Note that we rescaled the number density by multiplying by
the ratio of the total number of the original sample to that of the
bias-corrected sample. The purple-solid line is the boundary between the
thin- and thick-disk components, derived from the bias-corrected sample,
while the black-solid line in this panel is the original (biased) sample. 
Note that their difference is negligible.}
\label{fig:4}
\medskip
\end{figure*}

In this paper, we are most interested in the stellar populations of the
Galactic disk. As the typical halo star has low metallicity, slower
orbital rotation velocity, and higher $|W|$ velocity dispersion than its
thick-disk counterparts, we further culled the sample by including only
stars with \feh\ $>$ --1.2, \z\ $<$ 3 kpc,  distance of $d$ $<$ 4 kpc, 
$|W| <$ 100 \kms, and rotation velocity of \vph $>$ 50 \kms, which will serve to minimize
contamination from halo stars. We also restricted the sample to 
a range of $7 < R < 11$ kpc along the disk plane, as 
our program stars predominantly reside in this region.

In summary, our final sample satisfies the following conditions: 
0.45 $< (g - r)_0 <$ 0.80 and 13.5 $< r_0 <$ 20.5, 4500 $<$ \teff\ $<$ 6000 K, 
\logg\ $>$ 3.8, S/N $>$ 20, \z\ $<$ 3 kpc, $d < $ 4 kpc, 7 $< R <$ 11 kpc,
\feh\ $>$ --1.2, \vph\ $>$ 50 \kms, $|W| <$ 100 \kms\, for stars within
2$\sigma$ in the color-temperature relation of Figure \ref{fig:1}.
Figure \ref{fig:2} displays the spatial distribution of the final sample
of 23,908 stars as black dots. The red dot at ($R$,$Z$) = (8.0,0.025) kpc
is the Sun's location, and a blue-dashed line represents $Z$ = 0 kpc.

\subsection{Chemical Separation of the Thin- and Thick-disk Populations} \label{subsec:separation}

We chemically divide our program stars into the thin- and thick-disk
populations based on their $\alpha$-abundance ratios (as a function of
[Fe/H], following the similar way by \citealt{Lee11b}, as described
below). Figure \ref{fig:3} shows histograms of the $\alpha$-abundance
ratios over different cuts in [Fe/H]. With the exception of the lowest
and highest metallicity bins, shown in the upper-left panel and
lower-right panels, respectively, histograms for the remaining stars
were split using steps of 0.1 dex, listed at the top of each panel.
Then, assuming that each histogram consists of two unique stellar
populations contributed by the thin disk and thick disk, we fitted
Gaussians to each distribution. In this process, as the 
$\alpha$-distribution of the thick-disk component 
is rather broad in the range of --0.4 $<$ [Fe/H] $\leq$ --0.1, we first
fitted a Gaussian to the thin-disk component to obtain its best-fit
parameters; after fixing these, we derived the best-fit Gaussian for the
thick-disk population. As there is no apparent
contribution from the thin disk for the low-metallicity region (\feh\
$<$ --0.7) or from the thick disk for the highest metallicity region
(\feh\ $> +0.1$), we only fitted single Gaussian. The orange curves
represent the best-fit lines for each component obtained from the
Gaussian fits. Note that we have explicitly not included any presumed
contribution from the MWTD, which is shown as a distinct population from
the canonical thick disk by recent studies \citep{Carollo19,An20}.

We then identified boundary points (shown as the gray-solid lines in
Figure \ref{fig:3}) for the distributions in each metallicity bin. To
accomplish this for the highest and two lowest metallicity cuts, 
where single Gaussian exists, we determined the \afe\ value located 
3$\sigma$ away from the mean. As a result, we obtained a first estimate of the
division between the thin- and thick-disk populations. We repeated this
exercise by shifting the metallicity range used in Figure \ref{fig:3} by
$+$0.05 dex, and obtained another set of boundary positions between the
thin- and thick-disk populations. We then applied a linear least-square
fit to the two sets of boundary points as a function of \feh\ to derive
a dividing line of the two stellar populations. The resulting functional
form is \afe\ = --0.141 $\times$ \feh\ $+$ 0.140 over the metallicity
range of --0.65 $<$ \feh\ $<$ --0.25; beyond this, we adopted constant
values of \afe\ to $+$0.232 for [Fe/H] $<$ --0.65, and $+0.175$ for
[Fe/H] $>$ --0.25, respectively. Note that, for this exercise, we only
considered stars with spectra having S/N $>$ 50 (hence the best \afe\
and \feh\ determinations) in order to obtain the most reliable separation.

Our adopted chemical-dividing line between the two populations is shown
as a solid-diagonal line in the left panel of Figure~\ref{fig:4}, which
shows a logarithmic gray-scale plot of the number density of our dwarf sample
in the \feh\--\afe\ plane. Each bin has a size of 0.025 by 0.01 dex in
\feh\ and \afe, respectively and has at least two stars. The figure
clearly illustrates the two different populations as high-$\alpha$
(thick-disk) and low-$\alpha$ (thin-disk) sequences as a function of
\feh; the stars were assigned based on having distances from the solid
line of more than $\pm$0.05 dex (shown as dashed lines), in order to
accommodate the uncertainty in the derived \afe. Stars in the region 
between the dashed lines and the solid lines are not assigned to either
population. Through the application of this dividing scheme, we obtained
12,490 stars assigned to membership in the thin-disk population, and
6,712 stars assigned to the the thick-disk population, and use these
sub-samples below to carry out kinematic analysis.

\subsection{Potential Selection Biases}

Because the stars in our sample were selected for spectroscopic 
observations based on color and apparent magnitude, it may 
suffer from selection effects due to the target-selection algorithm 
used in SEGUE. For instance, in a magnitude-limited survey, because the 
metal-rich dwarfs are intrinsically brighter than the metal-poor dwarfs, there is 
a greater possibility to observe the metal-rich stars in a given field. On the 
other hand, the $(g-r)_{0}$ color cut may preferentially include the 
lower-mass metal-poor stars relative to the slightly higher-mass metal-rich 
stars, as the former could outnumber the latter, resulting 
in biasing our sample towards including more low-metallicity stars. If there exists
a significant metallicity bias in the SEGUE target selection, this may
impact the dividing criterion  we applied in Section \ref{subsec:separation}, and
our interpretations of the kinematic properties of the different components. 
Hence, we have carried out an exercise to test the sensitivity of our dividing 
scheme by these potential biases.

First, we obtained the selection function for our G- and K-type dwarfs 
by adopting the similar methodology that other studies 
used (e.g., \citealt{Schlesinger12,Nandakumar17,
Wojno17,Chen18, Mints19}). The basic idea of the method is to derive 
the number ratio of the spectroscopically targeted objects
to the photometrically available ones in a small bin on a color-magnitude
diagram (CMD) for each SDSS plug-plate. In this process, we considered a bin size
of 0.05 mag by 0.2 mag in color and magnitude, respectively, for
a CMD of ($g-r$)$_{\rm 0}$ and $r_{\rm 0}$. The selection function is 
regarded as the ratio of the number of stars selected for spectroscopic 
observation to the number of stars present in the direction of a given 
plug-plate with available photometry in each color and magnitude bin. 
After the calculations of the selection function, we corrected the 
selection bias of our dwarfs by multiplying the inverse of the selection 
function for each object.

Afterwards, we followed the exact same procedure described in 
Section \ref{subsec:separation} to derive the dividing line between the 
thin- and thick-disk populations. The right panel of Figure \ref{fig:4} 
shows the density map of the selection bias-corrected sample. The purple 
line denotes the boundary between the thin and thick disk, which
was derived from the sample corrected for the selection bias, while the
black-solid line from the original sample, which is not corrected for the selection bias.
We notice the negligible difference between the two lines in the figure. Because
we are not attempting to determine a precise metallicity distribution function 
from our sample, we decided to just use the original dwarf sample without the bias 
correction and the disk-separating criterion derived from the biased sample in 
the following analysis. This is also justified by the fact that because our sample was 
selected purely based on color and magnitude, it does not suffer from kinematic bias. 
Hence, any kinematic and dynamical properties that we are seeking to will not be 
greatly influenced by any potential selection bias caused by the color and magnitude.

\begin{figure*}
\centering
\includegraphics[width=1.0\linewidth]{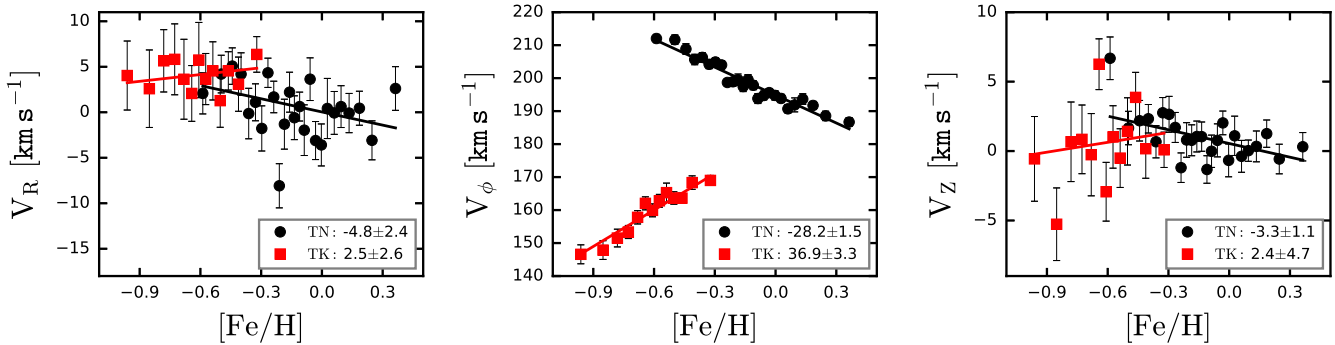}
\caption{Trend of median values of \vrd\ (left), \vph\ (middle), and \vz\ (right)
velocity components over metallicity. The black dots are for the thin-disk stars,
and the red squares for the thick-disk stars. Each symbol represents a median value
of 500 stars and error bars are calculated by bootstrapping of the sample in each symbol 1000 times.
The two-solid lines, which show the velocity gradients are obtained by
the least square fit to the median values. The slope of each velocity component for each
population is denoted with its associated uncertainty in the square box.}
\label{fig:5}
\end{figure*}

\begin{table}
\caption{Summary of Kinematics for Thin- and Thick-Disk Populations}
\begin{center}
\begin{tabular}{crr}
\hline
\hline
\multicolumn{3}{c}{Gradients with Metallicity (\kmsdex)} \\
\hline
Component & Thin disk & Thick disk \\

\hline
$\partial V_{\rm R}$/$\partial$\feh\    &  --4.8 $\pm$ 2.4 &   2.5 $\pm$ 2.6\\
$\partial V_{\rm \phi}$/$\partial$\feh\ & --28.2 $\pm$ 1.5 &  36.9 $\pm$ 3.3\\
$\partial V_{\rm Z}$/$\partial$\feh\    &  --3.3 $\pm$ 1.1 &   2.4 $\pm$ 4.7\\
\hline
$\partial \sigma_{V_{\rm R}}$/$\partial$\feh\    &   0.6 $\pm$ 1.2 & --43.2 $\pm$ 4.6 \\
$\partial \sigma_{V_{\rm \phi}}$/$\partial$\feh\ &   1.9 $\pm$ 0.6 & --17.4 $\pm$ 2.6 \\
$\partial \sigma_{V_{\rm Z}}$/$\partial$\feh\    & --6.0 $\pm$ 1.2 & --20.6 $\pm$ 1.4 \\
\hline
& & \\

\hline
\hline
\multicolumn{3}{c}{Mean Velocities and Velocity Dispersions (\kms)} \\
\hline
Component & Thin disk & Thick disk \\

\hline
$\langle V_{\rm R} \rangle$    &   0.1 $\pm\ 0.4$ & $  3.8\ \pm\ 0.8$ \\
$\langle V_{\rm \phi} \rangle$ & 197.2 $\pm\ 0.2$ & $159.2\ \pm\ 0.5$ \\
$\langle V_{\rm Z} \rangle$    &   0.9 $\pm\ 0.2$ & $  0.6\ \pm\ 0.5$ \\
\hline
$\langle\sigma _{V_{\rm R}}\rangle$    & $40.2\ \pm\ 0.3$ & $62.9\ \pm\ 0.5$ \\

$\langle\sigma _{V_{\rm \phi}}\rangle$ & $26.0\ \pm\ 0.2$ & $39.1\ \pm\ 0.3$ \\
$\langle\sigma _{V_{\rm Z}}\rangle$    & $21.3\ \pm\ 0.2$ & $38.1\ \pm\ 0.3$ \\
\hline

\end{tabular}
\tablecomments{The errors in the mean and dispersion are based on 1000 bootstrap
re-samples. The mean of the velocity and dispersion is calculated
by taking an average of the symbols for each population in Figures \ref{fig:5} and \ref{fig:6},
respectively.}
\end{center}
\label{tab:1}
\end{table}

\section{Kinematic Properties of the Thin- and the Thick-disk Populations}\label{sec:result_vel}
\subsection{Correlation of Rotation Velocity With Metallicity}

Figure \ref{fig:5} shows the distributions of the \vrd, \vph, and \vz\
velocity components, as a function of \feh, for the thin-disk (black dots)
and thick-disk (red squares) populations. Each symbol represents a median
value of 500 stars in a sample sorted in ascending order of [Fe/H]. The
error bar is a standard deviation obtained from bootstrapping of the
sample 1000 times. The two solid lines representing the velocity
gradients are obtained by a least-square fit to the median values of
each population. The derived slope of each velocity component for each
population is indicated in the legend of each panel. We also
performed a least-square fit to the entire sample for each velocity
component, and found no discernible discrepancy within the derived
uncertainties of the gradients, compared to those in Figure
\ref{fig:5}.

From inspection of Figure~\ref{fig:5}, there exist only small gradients
for \vrd\ and \vz\ velocities with metallicity for both populations, as
found in previous studies (e.g., \citealt{Guiglion15,Wojno16}), 
even though the \vz\ gradient for the thick disk has a relatively 
larger uncertainty. By contrast, the \vph\
gradients for the thick and thin disks are highly significant, and of
opposite signs: $+$36.9 $\pm 3.3$ \kmsdex\ and --28.2 $\pm$ 1.5 \kmsdex,
respectively. Moreover, the mean of the \vph\ component of the thin disk
($+$197.2 $\pm$ 0.2 \kms) is significantly larger than that of the thick
disk ($+$159.2 $\pm$ 0.5 \kms), a collective lag in the rotational
velocity of the thick disk that has been reported by numerous previous
authors (e.g., \citealt{Guiglion15,Rojas16, Belokurov20}).

We recall results on \vph\ gradients with [Fe/H] reported by
a number of previous studies.  From a sample of SEGUE G dwarfs with less-precise 
distances and proper motions, \citet{Lee11b} reported a \vph\ gradient of 
$+45.8 \pm 2.9$ \kmsdex\ and --22.3 $\pm 1.6$ \kmsdex\ and for 
the thick and thin disk, respectively. \citet{Adibekyan13} examined 
F-, G-, and K-type dwarfs from the HARPS
sample \citep{Mayor03, LoCurto10, Santos11}, and derived \vph\ 
gradients of $+$41.9 $\pm$ 18.1 \kmsdex\ and --16.8
$\pm$ 3.7 \kmsdex\ for the thick disk and thin disk, respectively.
\citet{Blanco14}, using \gaia\ ESO F-, G-, and K-type stars, reported
values for the \vph\ slopes of $+$43 $\pm$ 13 \kmsdex\ and --17.6 $\pm$
6 \kmsdex\, for the thick and thin disk, respectively.  The studies by 
\citet{Guiglion15} and \citet{Wojno16} reported a \vph\
gradient of $+$49 $\pm$ 10 \kmsdex\ and $+$51 $\pm$ 10 \kmsdex,
respectively, for the thick-disk population, which agree with our values
within the errors. As for the thin disk, \citet{Wojno16} found --11 $\pm$ 1 \kmsdex, 
whereas \citet{Guiglion15} claimed a positive rotation 
velocity gradient. They argued that this is because their
sample covers a larger metallicity range (--1.0 $<$ \feh\ $<$ $+$0.5) 
than other studies. They indeed found a slightly negative gradient of --5 $\pm$
5 \kmsdex\ by narrowing the metallicity range (--0.7 $<$ \feh\ $<$ $+$0.2),
but still less steep than others. From giants observed by LAMOST with
$Gaia$ DR2 astrometry, \citet{Yan19} found a positive slope ($+$30.87 $\pm$
0.001 \kmsdex) and negative slope (--17.03 $\pm$ 0.001 \kmsdex) for the
thick and thin disk, respectively. Interestingly, their slope of the gradient in the
thick disk is much shallower than found by other studies.  

\citet{ReFiorentin19} carried out a thorough investigation
of the rotation velocity gradient with metallicity at various distances
from the Galactic center, using 58,882 red giants from APOGEE DR14
(\citealt{Abolfathi18}) and proper motions from \gaia\ DR2. For the
thick-disk population, they derived slopes between $+$20.7 and $+$60.1
\kmsdex, increasing with distance from the Galactic center. At the Solar
radius ($R$ $\sim$ 8 kpc), their reported gradient was $+$43.5 \kmsdex.
They consistently found negative gradients for the thin-disk population
over 5 $< R <$ 13 kpc, in the range --43.8 to --16.3 \kmsdex, the lowest at the 
Solar radius. 

We have carried out a similar exercise. After dividing our sample into 
three regions: 7 $< R \leq$ 8 kpc, 8 $< R \leq$ 9 kpc, and 9 $< R \leq$ 11 kpc, 
we derived gradients of --34.0 $\pm$ 2.4 \kmsdex, --27.6 $\pm$ 1.2 \kmsdex, and 
--28.4 $\pm$ 1.8 \kmsdex, respectively, for the thin disk, and  
27.7 $\pm$ 4.0 \kmsdex, 42.1 $\pm$ 3.5 \kmsdex, and 61.2 $\pm$ 7.6 \kmsdex, 
respectively, for the thick disk. Even though the magnitudes of the gradients 
are slightly different, the overall trends with Galactocentric distance 
are the same as found by \citet{ReFiorentin19}.   

To summarize, our derived \vph\ gradient ($+$36.9 \kmsdex) 
for the thick disk generally agrees with those from other studies mentioned 
above. However, it appears that our \vph\ gradient (--28.2 \kmsdex) of the 
thin disk is slightly steeper, by about --10 \kmsdex, than other studies. 
The main source of the difference is the spatial coverage of the thin-disk 
stars considered. While our chemically separated thin-disk stars include many 
objects (about 41\%) in the region \z\ $>$ 0.8 kpc, the thin-disk samples in 
the aforementioned studies consist of stars predominantly in the 
region \z\ $<$ 0.8 kpc. Indeed, when we restrict our thin-disk stars to the 
region \z\ $<$ 0.8 kpc, we obtain a less-steep gradient, of about --25 \kmsdex.
The reason our sample includes many thin-disk 
stars with \z\ $>$ 0.8 kpc is likely due to the combination of the 
sampling region in the SEGUE survey and its target-selection strategy. 
The bright limit of the SDSS photometry hinders the 
observation of (bright) stars close to the Galactic plane.

\begin{figure*}
\centering
\includegraphics[width=1.0\linewidth]{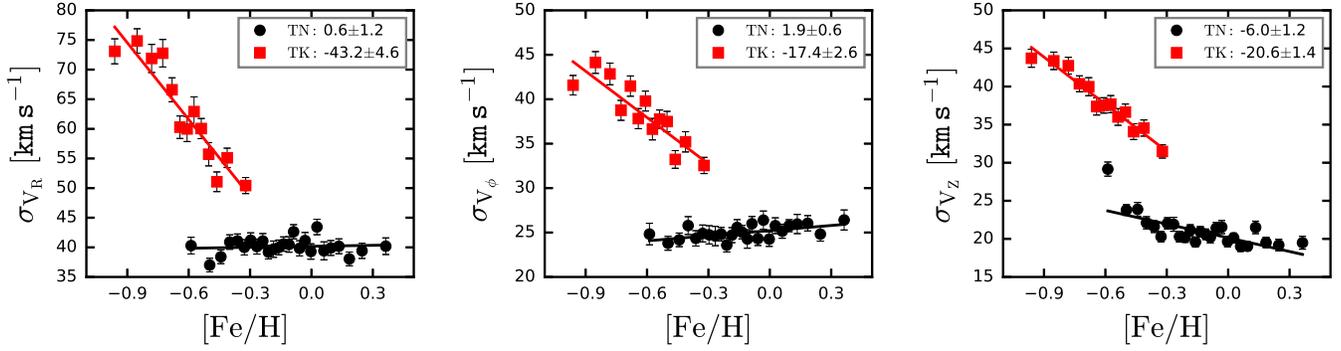}
\caption{Same as in Figure \ref{fig:5}, but for \vrd, \vph, and \vz\ velocity dispersions.}
\label{fig:6}
\end{figure*}

\begin{figure*}
\centering
\includegraphics[width=1.0\linewidth]{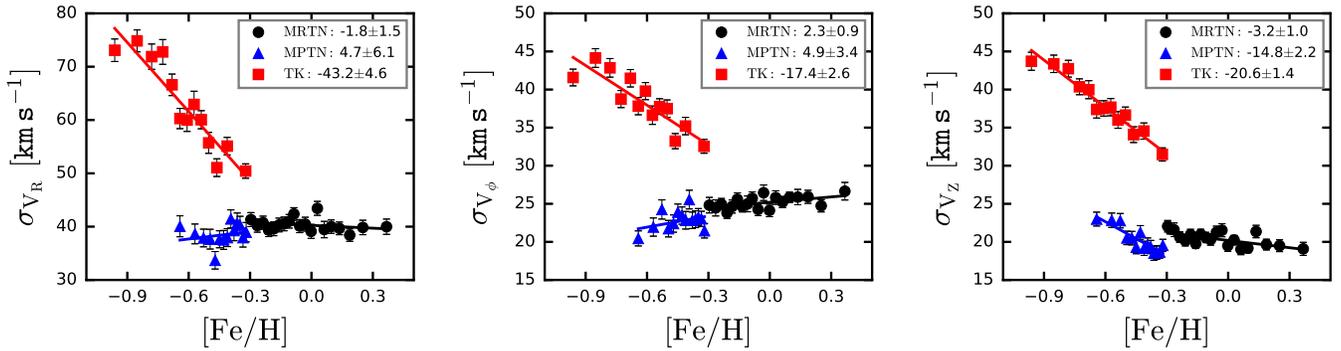}
\caption{Same as in Figure \ref{fig:5}, but the thin-disk population is divided into metal-rich (black dots;
[Fe/H] $>$ --0.3) and metal-poor (blue triangles; [Fe/H] $\leq$ --0.3)
groups, indicated in the legends as MRTN and MPTN, respectively. The
metal-poor group only includes stars with thin-disk kinematics (see text
for more details). The blue triangles are the median value of 200 stars,
while the black dots are a median value for 500 stars.}
\label{fig:7}
\end{figure*}

\begin{table}
\caption{Kinematic Parameters used for Membership Probability}
\begin{center}
\begin{tabular}{ccc}
\hline
\hline
Parameter & Thin & Thick \\
\hline
X & 0.94 & 0.06 \\
$\sigma_{U}$ (\kms) & 43 & 67 \\
$\sigma_{V}$ (\kms) & 28 & 51 \\
$\sigma_{W}$ (\kms) & 17 & 42 \\
$V_{\rm asy}$ (\kms) & --9 & --48 \\
\hline
\end{tabular}
\tablecomments{X represents the fraction of thin- or thick-disk stars, and $V_{\rm asy}$ is the asymmetric
drift relative to the LSR.}
\end{center}
\label{tab:2}
\end{table}

\begin{table*}
\caption{Mean Values of Velocities, Velocity Dispersions, and \afe\ for MRTN and MPTN Stars}
\begin{center}
\begin{tabular}{ccccccccc}
\hline
\hline 
& $N$ & $\langle$$V_{\rm R}$$\rangle$ & $\langle$$V_{\rm \phi}$$\rangle$ & $\langle$$V_{\rm Z}$$\rangle$ & $\langle\sigma_{V_{\rm R}}\rangle$ & $\langle\sigma_{V_{\rm \phi}}\rangle$ & $\langle\sigma_{V_{\rm Z}}\rangle$ &  $\langle$\afe$\rangle$ \\
 & & (\kms) & (\kms) & (\kms) &(\kms) & (\kms) & (\kms) &  \\
\hline
MRTN & 9466 & --0.7 & 194.1 & 0.6 & 40.4 & 25.6 & 20.4 & 0.03 \\
MPTN & 2855 &   2.9 & 208.2 & 2.2 & 38.7 & 23.1 & 20.4 & 0.11 \\
\hline
\end{tabular}
\tablecomments{$N$ is the number of stars. Possible contaminating
thick-disk stars have been excluded (see text for details).}
\end{center}
\label{tab:3}
\end{table*}

\subsection{Velocity Dispersion Gradients with Metallicity} \label{subsubsec:velo-dispersion}

Figure \ref{fig:6} exhibits the trends of \vrd, \vph, and \vz\ velocity
dispersions with metallicity. Following the same procedure as in Figure
\ref{fig:5}, we derived the slope of the velocity dispersion for each
velocity component; the same symbols are used as in Figure \ref{fig:5}.
The figure clearly shows that the chemically selected thick-disk
population exhibits higher velocity dispersions and steeper gradients
than the thin-disk population for all velocity components. Table
\ref{tab:1} summarizes the gradient of each velocity component and
velocity dispersion, and their mean values, calculated by averaging the
median points for each population in Figures \ref{fig:5} and
\ref{fig:6}. From inspection of the table, the \vrd\ component exhibits
the largest velocity dispersion and the steepest dispersion gradient among the
three velocity components of the thick disk. 

One notable feature in Figure \ref{fig:6} is that, unlike the velocity
gradient, a negative slope is found for $\sigma_{V_{\rm R}}$,
$\sigma_{V_{\rm \phi}}$, and $\sigma_{V_{\rm Z}}$ for the thick-disk
population, as a function of metallicity, whereas the thin-disk stars
exhibit an almost flat velocity dispersion gradient for all three
velocity components. However, a clear increase of $\sigma_{V_{\rm
Z}}$ with decreasing metallicity is seen for the more metal-poor (\feh\
$<$ --0.3) thin-disk stars in the metal-poor region (right panel
of Figure~\ref{fig:6}). The dispersion at [Fe/H] = --0.6 is as high as
that of the thick disk at [Fe/H] = --0.3. 

Due to the results above, we investigated the hotter kinematics of the 
metal-poor thin-disk stars (MPTN; [Fe/H] $<$ --0.3) if these could arise
because of contamination from thick-disk stars, following the method
described by \citet{Bensby03, Bensby14}. The basic idea is that, by
assuming that stars of the thin- and thick-disk populations have
Gaussian distributions with different space velocities ($U$, $V$, and
$W$) and asymmetric drift, stars can be separated by their probability
of belonging to one or the other population. We employed the parameters
listed in Table \ref{tab:2}, which are adopted from
\citet{Reddy06}. In the table, $X$ is the fraction of stars that belong
to each disk component in the local sample of stars. If a star has a
higher likelihood of being a member of the thick disk, relative to the
thin disk, that is, Pr(TK)/Pr(TN) $> 0.5$, this star is assigned to the
thick disk. On the other hand, thin-disk membership is assigned if a
star has Pr(TK)/Pr(TN) $<$ 0.5. We applied this procedure to the MPTN
stars, and found that among the 3,024 chemically separated MPTN stars,
2,855 stars are kinematically assigned to the thin disk, while the
number of the stars assigned to the thick disk is only 169, suggesting
that contamination from the thick-disk stars in the thin-disk subsample
is minimal.

\begin{figure*}
\centering
\plotone{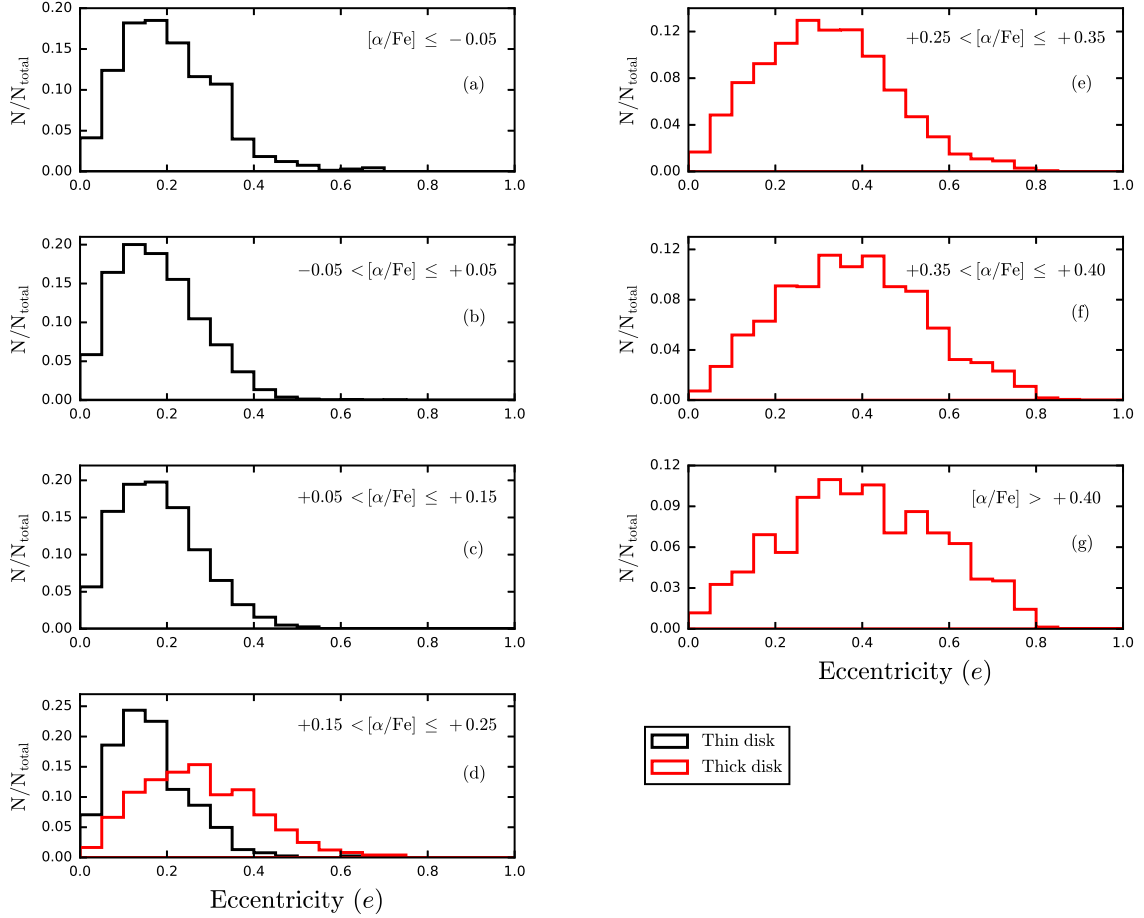}
\caption{Histograms of eccentricities for the thin-disk (black line) and the thick-disk (red line) stars in
different \afe\ bins, as indicated in the legends of each panel. The \afe\ range increases from (a) to (g).}
\label{fig:8}
\medskip
\end{figure*}

From the kinematically identified metal-poor thin-disk stars with [Fe/H] $<$
--0.3, we have examined the distribution of the velocity dispersions
again, as shown in Figure \ref{fig:7}. The MPTN stars are represented
with blue triangles; each triangle corresponds to 200 stars. We can see that the \vz\
velocity dispersion of the MPTN stars increases with decreasing
metallicity, and its trend is similar to that of the MPTN stars in Figure
\ref{fig:6}, whereas the \vrd\ and \vph\ velocity dispersions exhibit
almost no gradients. By comparison, the \vrd\ and \vz\ dispersion
gradients of the metal-rich thin-disk (MRTN; [Fe/H] $>$ --0.3) stars are
almost flat, while the slope of the \vph\ component exhibits a small
positive value. These stars are represented by black dots in Figure
\ref{fig:7}; each black dot corresponds to 500 stars.

Table \ref{tab:3} lists the total numbers and mean values of the
velocity components, their velocity dispersions, and \afe\ for the MRTN
and MPTN stars. Note that the MPTN stars do not include the objects with
probable thick-disk kinematics rejected as above. The dispersions of the
\vrd\ and \vz\ components of the MPTN stars are nearly the same as the
ones from the entire thin-disk stars (see Table \ref{tab:1}), but 
$\sigma_{V_{\rm \phi}}$ is slightly smaller. We also observe that 
$\langle$\vph$\rangle$ becomes a little bit larger, while the change 
in $\langle$\vrd$\rangle$ and $\langle$\vz$\rangle$ is very small. We 
conclude from these results that the small possible contamination 
from stars with thick-disk kinematics in the MPTN
sub-sample has little effect on our results. Thus, our chemical criterion for
thin- and thick-disk separation appears quite robust, and suitable for
investigations of the kinematics of each population. In this regard, it
is worth mentioning that \citet{Rojas16} distinguished thin-disk stars
into three groups: metal-rich, metal-intermediate (MI), and metal-poor
(MP) stars. The MI and MP regions are divided at \feh\ $\sim$ --0.25,
which is close to our adopted dividing value. They investigated the
dispersion of rotation velocity for each population, and identified a
higher dispersion for the MPTN stars than for the MITN and MRTN
populations. Furthermore, a close examination of the right panel 
of Figure \ref{fig:7} clearly indicates some offset in $\sigma_{V_{\rm Z}}$ 
between the MRTN and MPTN samples. This may support the claim of an independent 
evolution of MPTN stars by \citet{Rojas16}.

Comparing with other studies, \citet{Wojno16} found velocity dispersions
of ($\sigma_{V_{\rm R}}$, $\sigma_{V_{\rm \phi}}$, $\sigma_{V_{\rm Z}}$)
= (32, 19, 15) \kms\ in cylindrical coordinates for the thin disk, and
($\sigma_{V_{\rm R}}$, $\sigma_{V_{\rm \phi}}$, $\sigma_{V_{\rm Z}}$) =
(49, 36, 29) \kms\ for the thick disk. These values are slightly lower
than ours, as listed in Table \ref{tab:1}. Interestingly,
\citet{Blanco14} reported no significant slope in $\sigma_{V_{\rm
\phi}}$ and $\sigma_{V_{\rm Z}}$ for the two disk populations, and found
similar rotation velocity dispersions (about 45 \kms) for both the thin
disk and thick disks at \feh\ $\sim$ --0.3. \citet{Li17} derived a
vertical velocity dispersion ($\sigma_{V_{\rm Z}}$) of 58.8 $\pm$ 0.3
\kmsdex\ for the thick disk from 2035 giants observed by LAMOST, a value
much larger than reported by the other studies (including ours).
\citet{Hayden20} investigated about 62,000 stars from GALAH, and
calculated a vertical velocity dispersion of about 8 \kms\ at \afe\ = 0
and [Fe/H] = 0, increasing to over 50 \kms\ for more
$\alpha$-enhanced, metal-poor stars. \citet{Belokurov20} obtained
($\sigma_{V_{\rm R}}$, $\sigma_{V_{\rm \phi}}$, $\sigma_{V_{\rm Z}}$) =
(31 $\pm$ 6, 21 $\pm$ 4, 26 $\pm$ 5) \kms\ and ($\sigma_{V_{\rm R}}$,
$\sigma_{V_{\rm \phi}}$, $\sigma_{V_{\rm Z}}$) = (73 $\pm$ 7, 47 $\pm$
6, 48 $\pm$ 6) \kms\ for the thin- and thick-disk stars, respectively,
in the ranges of 2 $<$ \z\ $<$ 3 kpc and --0.7 $<$ \feh\ $<$ --0.2. The
dispersions of their thin-disk population generally agree well with
ours, taking into account the quoted uncertainties, but those of 
their thick-disk population are slightly higher than ours.   

Summarizing, the different trends of the velocities and velocity
dispersions over [Fe/H] between the thin-and thick-disk populations
reported here and elsewhere clearly imply that the two populations had
different star-formation and/or evolutionary histories. In particular, 
the negative slope of the rotation velocity for the thin disk may indicate
that the radial mixing plays an important role in shaping its current
form. 

\begin{figure}
\centering
\plotone{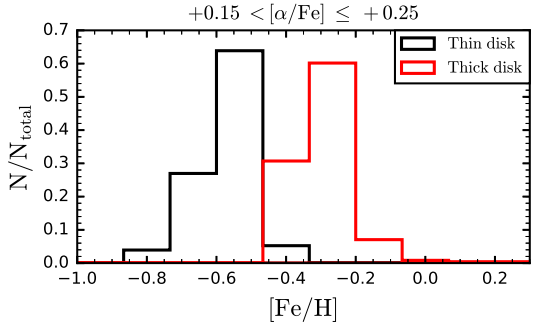}
\caption{Metallicity distribution functions (MDFs) of the thin-disk (black line) and thick-disk (red line) 
stars in the range of $+$0.15 $<$ \afe\ $\leq$ $+$0.25.}
\label{fig:9}
\medskip
\end{figure}

\begin{figure*}
\centering
\plotone{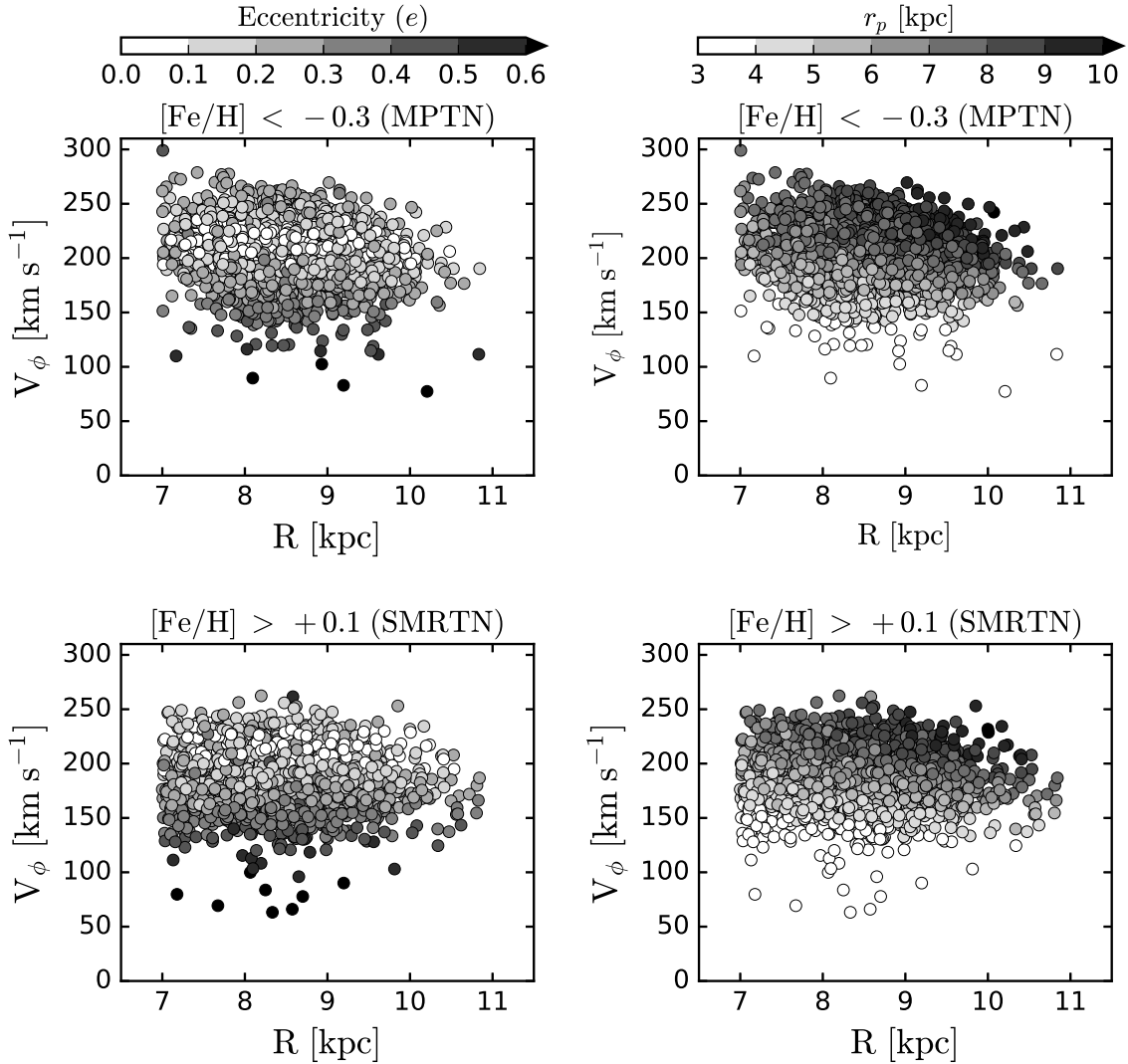}
\caption{Left~panels: Gray-scale distribution of eccentricities of metal-poor (upper panel; \feh\ $<$ --0.3)
and super metal-rich (bottom panel; \feh\ $>$ $+$0.1) thin-disk stars in the $R$--\vph\ plane. $R$
is the Galactocentric distance projected onto the Galactic plane. The gray-scale bar represents
the scale of the eccentricity. Right~panels: Same as in the left panels, but for the perigalactic
distance ($r_{\rm p}$).}
\label{fig:10}
\medskip
\end{figure*}

\begin{figure*}
\centering
\plotone{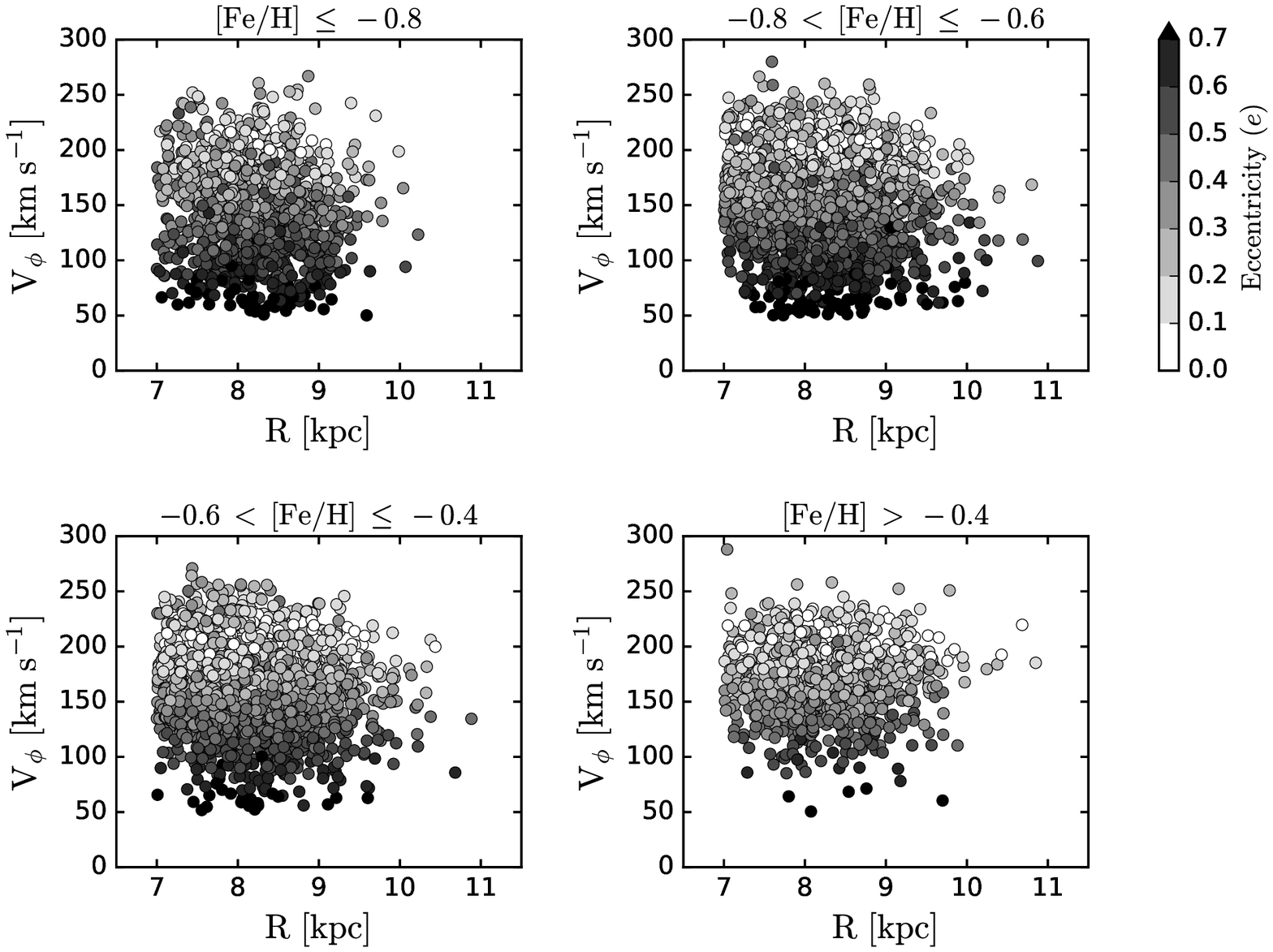}
\caption{Same as in the left panels of Figure \ref{fig:10}, but for the thick-disk stars in four 
metallicity ranges.}
\label{fig:11}
\medskip
\end{figure*}

\section{Dynamical properties of the thin and thick disks}\label{sec:result_orbit}

\subsection{Eccentricity Distribution Over \afe}

Among the available stellar dynamical properties, the
eccentricity distribution of a given population is a very useful
tracer of its origin and evolution. For example, \citet{Sales09}
investigated the distribution of the eccentricities of thick-disk
stars simulated from four different thick-disk formation scenarios, and
reported that the orbital eccentricity distribution of each model
exhibits distinctly different properties. Thus, in order to characterize
the likely formation history of our chemically separated thick-disk
population, we have considered the eccentricities of the stars over
different ranges of \afe, as shown in Figure \ref{fig:8}. In this figure,
the range of \afe\ is indicated in the legend at the top right of each
panel, and increases from the top-left to the bottom-right panel (e.g.,
from (a) to (g)).

Inspection of Figure \ref{fig:8} reveals that the thin-disk
population (black line in the figure) has a skewed distribution up 
to relatively high eccentricity, $e \sim  0.5$, with a peak at a value
$e <$ 0.2. It is also interesting to note that the width of the
distribution becomes narrower as one moves toward higher values of \afe.
In other words, the fractions of high $e$-stars decrease with increasing
\afe. On the other hand, the thick disk (red line) exhibits a much broader
distribution of orbital eccentricities, with an extended tail to much
higher $e$. The $e$-distribution of the thick disk becomes broader with
increasing \afe, and there are greater fractions of high-$e$ stars,
resulting in a shift of the peak to higher $e$. Close inspection at the
overlapping region at $+$0.15 $<$ \afe\ $<$ $+$0.25 indicates that the
distributions of the two populations appear very different. The
peak of the eccentricity distribution of the thick disk clearly occurs
at higher $e$ than that of the thin disk in this range of \afe. 

In order to examine the region of $+$0.15 $<$ \afe\ $<$ $+$0.25 
more closely, we have considered the MDFs of the thin- and thick-disk
populations (see Figure \ref{fig:9}). The black histogram 
is for the thin disk, while the red histogram is for the thick disk. The two 
populations clearly exhibit very different MDFs. The MDF of the metal-poor 
thin-disk population peaks at [Fe/H] = --0.55, compared to [Fe/H] = --0.25 
for the metal-rich thick-disk population. The width of 
each distribution is almost the same, however. This suggests that they have 
experienced different chemical-evolution histories, even though 
they share similar $\alpha$-abundance ratios. 

The general properties of our derived eccentricity distributions for
the two disks qualitatively agree with previous studies. For
instance, \citet{Hayden18} studied the Galactic disk with 3,000 stars
from the $Gaia$-ESO survey, and reported not only the increasing trend of
eccentricity with increasing [Mg/Fe], but also a median eccentricity of
$e \sim$ 0.33 for the high-[Mg/Fe], metal-poor stars, which is
very similar to our value (0.4) of the peak eccentricity for \afe\ $>$
$+$0.40. Using giants observed by LAMOST with available \gaia\ DR2 astrometry,
\citet{Yan19} found an increasing trend of orbital eccentricities with
decreasing metallicity for the thick-disk population. We can infer 
this trend in Figure \ref{fig:8} as well, because \afe\ generally 
increases with decreasing [Fe/H], as can be seen in Figure \ref{fig:4}.

According to the study by \citet{Sales09}, the eccentricity distribution
of thick-disk stars formed through heating by minor mergers exhibits a
secondary peak. However, since our results in Figure \ref{fig:8} do not
appear to have a secondary peak for all \afe\ bins, we can rule out
this scenario. Note, however, that the existence 
of the high eccentricity secondary peak may depend on the initial 
conditions of the interacting dwarf galaxies in the simulation.
On the other hand, the eccentricity distribution of the 
accretion model seems similar to that of panel (g) of Figure \ref{fig:8}, 
even though the peak of panel (g) is lower than the prediction of the model. 
Nevertheless, it cannot explain the change in the distribution with \afe, and 
the lack of stars with significantly high eccentricity stars in this panel. 
Consequently, we can discard this scenario as a mechanism for the thick-disk 
formation as well. Considering the migration and merger models, it is hard to tell
quantitatively from the eccentricity distribution which one is a better
prescription for thick-disk formation, as they predict qualitatively
very similar $e$-distribution to our results. To distinguish one from
the other, we consider the investigation described below.

\subsection{Distribution of Eccentricities and Perigalactic Distances 
in the $R$--\vph\ Plane} \label{sec:ecc}

In this section, we examine the orbital properties of the MPTN ([Fe/H]
$<$ --0.3), super metal-rich ([Fe/H] $>$ $+$0.1) thin-disk (SMRTN), and
the thick-disk populations in greater detail. Figure~\ref{fig:10}
compares how the eccentricity (left panels) and perigalatic distance
distributions (right panels) differ, in the $R$--\vph\ plane, between
the MP (upper panels) and SMR (lower panels) populations of the thin
disk. In the upper-left panel of Figure~\ref{fig:10}, for the MPTN
stars, we observe a group of stars (hereafter, G1) with high
eccentricity ($e >$ 0.2) and low orbital rotation velocity (\vph\ $<$
170 \kms). Stars with similar properties to G1 can also be seen
in the bottom-left panel for the SMR stars. One clear difference for
stars with $e > 0.2$ between the MP and SMR stars is that almost all of
the SMRTN stars have \vph\ $<$ 180 \kms, while there are also high-\vph\
stars ($ > 240$ \kms) in the MPTN population. The stars with $e <$ 0.1
are mostly concentrated between \vph\ = 190 and 220 \kms\ in both populations.   

The upper-right panel of Figure \ref{fig:10} indicates, as expected, 
that the \rperi\ distance correlates with $R$, such that as the Galactocentric 
distance increases, the perigalactic distance increases as well. In addition, 
there is a tendency for the \vph\ velocity to increase as the \rperi\ 
distance increases at a given $R$. It also appears that the MPTN stars show a negative 
correlation between \vph\ and $R$, whereas no such trend exists for the SMRTN 
stars (see the right panels of Figure \ref{fig:10}).

Quantitatively, we estimated that approximately 52\% of the stars with
\vph\ $>$ 240 \kms\ are on circular orbits ($e \leq$ 0.2) for the MPTN
population. These high-\vph\ stars have a median value
of \rperi\ $\sim$ 8.3 kpc and \rapo\ $\sim$ 11.5 kpc, suggesting that
most of them move around the Galactic center in the outer edge of the Solar circle.
\citet{Mikolaitis14} and \citet{Rojas16} also reported their MP
thin-disk stars are dominant in outer-disk region. Consequently, what
Figure \ref{fig:10} suggests is that the high-\vph\ velocity stars of the
MPTN population rotate faster in the outer region. On the other hand, 
due to their too small \rperi\ (median \rperi\ $\sim$ 4.7 kpc) and high
eccentricity, the G1 stars reach the solar circle from the inner-disk region. 

As far as the SMRTN population is concerned, the lower panels of Figure
\ref{fig:10} indicate that there are stars with high eccentricity, low
orbital rotation velocity, and small \rperi\ present. We derived that
52\% of the SMRTN stars have $e <$ 0.2, with a median
\rperi\ of 7.0 kpc. The stars with $e <$ 0.2 and \rperi\ $>$ 7 kpc account for 27\%
of the SMRTN population. A similar trend is reported by
\citet{Hayden18}, who showed that a quarter of their metal-rich
thin-disk stars have \rperi\ $>$ 7 kpc. On the other hand,
\citet{Kordopatis15} reported that half of super metal-rich
thin-disk stars in the Solar vicinity from RAVE DR4
\citep{Kordopatis13} have low eccentricity ($e\ \leq\ 0.15$).

One step further, we have examined the eccentricity distribution in the $R$--\vph\ plane
for the thick-disk stars, in a similar fashion as for the thin-disk
stars shown in the left panels of Figure~\ref{fig:10}. As can be seen in
Figure \ref{fig:11}, we have divided the sample into four metallicity
bins. Inspection of this figure clearly shows that the fraction of
high-eccentricity stars is larger than for the thin-disk 
population (this is also obvious in Figure \ref{fig:8}).
Unlike the thin-disk population, however, the high-$e$ ($>$ 0.5) stars
are mostly concentrated in the region of \vph\ $<$ 120 \kms\ for all
four metallicity bins. The portion of the high-$e$ stars decreases with
increasing metallicity, indicating that the orbits of the more
metal-poor thick-disk stars tend to be more perturbed. We found that
about 77\% of the thick-disk stars have \rperi\ $<$ 6, with a median
value of \rperi\ $\sim$ 4.5 kpc.

\begin{figure}
\centering
\plotone{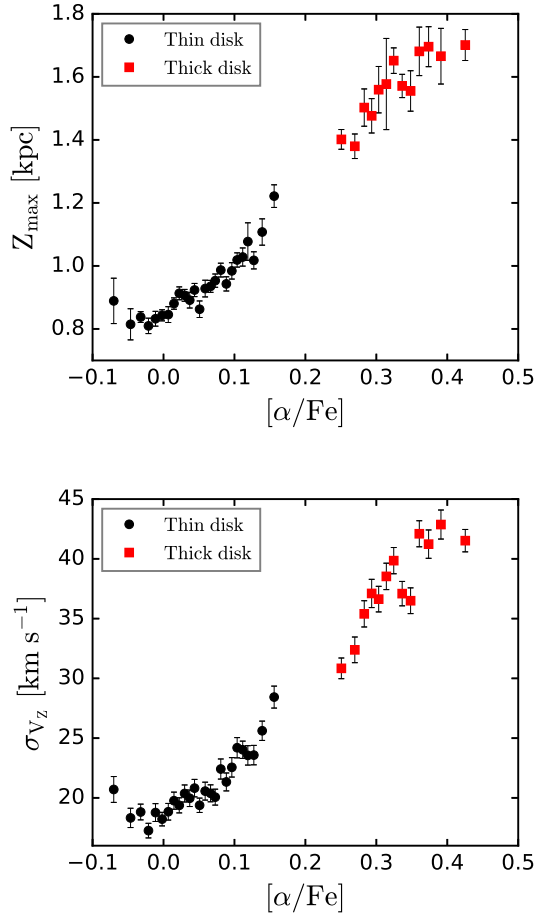}
\caption{Distributions of \zmax\ (top panel) and $\sigma_{V_{\rm Z}}$
(bottom panel), as functions of \afe. Each symbol represents the median 
of 500 stars; error bars are calculated from
1000 bootstrap resamples of the 500 stars.
The black dots and red squares represent the thin- and thick-disk stars, respectively.}
\label{fig:12}
\medskip
\end{figure}

\subsection{Trends of \zmax\ and $\sigma_{V_{\rm Z}}$ with \afe}\label{sec:333}

Finally, we investigated how the \vz\ velocity dispersion and maximum
vertical distance from the Galactic plane (\zmax) change with \afe.
Figure~\ref{fig:12} displays the trend of \zmax\ (top panel) and
$\sigma_{V_{\rm Z}}$ (bottom panel) as a function of \afe, for the thin-
(black dots) and thick- (red squares) disk stars. Each symbol represents
the median of 500 stars; error bars are calculated from 1000 bootstrap
resamples of the 500 stars. 

From inspection of the figure, both \zmax\ and $\sigma_{V_{\rm Z}}$ generally
rise with increasing \afe\ for both disk samples. This trend can be understood
in the sense that, because the high-$\alpha$ stars are generally old, they
have had more time to be perturbed to have higher \vz,
resulting in larger excursions from the Galactic plane. The high-$\alpha$
thin-disk stars reach as high as \zmax\ = 1.22 kpc.

This behavior has also been observed by other studies using different stellar
samples. For instance, \citet{Haywood13} obtained high-resolution
spectra of F-, G-, and K-type stars, and derived their ages. They
separated the thin- and thick-disk populations in the \feh\--\afe\ 
plane, and found that some of the thin-disk stars are older than 8 Gyrs, 
which overlaps with the younger thick-disk stars. Note that the age 
overlap between the thin- and thick-disk stars 
is also reported by observational and theoretical 
studies (\citealt{Delgado19,Rojas16,Spitoni19}). However, \citet{Spitoni19} 
asserted that the overlap in age might be caused by observational errors, because 
the uncertainty of asteroseismic ages used in their model tend to increase 
with increasing age.

\citet{Haywood13} reported that the oldest metal-poor thin-disk stars have high
\zmax, and that there is a significant increase in \zmax\ at near \afe\ $\sim$
$+$0.1, which can also be seen in the top panel of Figure \ref{fig:12}.
\citet{Rojas16} reported the similar behavior by analyzing the variation
of \z\ distance with [Mg/Fe] for disk-population stars. According to
their Figure 7, the \z\ distance of the MPTN population increases with
increasing [Mg/Fe], and the \z\ distance of the most [Mg/Fe]-rich stars
of the MP thin disk is comparable to that of the youngest (lowest
[Mg/Fe]) thick-disk stars. We also note
that there is continuous decrease of \zmax\ with declining \afe\ for the
thick-disk stars in the top panel of Figure \ref{fig:12}. \citet{Haywood13}
and \citet{Rojas16} found similar trends for the thick disk as well.
 
In the bottom panel of Figure \ref{fig:12}, the highest $\sigma_{V_{\rm
Z}}$ is around 28 \kms, in the most $\alpha$-rich bin among the
thin-disk population. \citet{Haywood13} also found an upward trend of
$\sigma_{V_{\rm Z}}$ with increasing \afe; for the thin-disk sequence,
their highest $\sigma_{V_{\rm Z}}$ $\sim$ 35 \kms, in good agreement
with ours. The thick-disk population exhibits a continuous increase
of $\sigma_{V_{\rm Z}}$ with increasing \afe, except for the downward trend
shown in the region of \afe\ $>$ $+$0.4, which is also reported by
\citet{Minchev14}. 

\section{Discussion on the Disk Formation and Evolution History}\label{sec:discussion}

It is clear from the different dynamical signatures observed in the
previous sections that the thin disk and thick disk have experienced
different formation and/or evolution histories. Before discussing these
formation histories in more detail, we first summarize the main properties
of the disk populations we have found:

\begin{enumerate}

\item Thin disk -- Consists of metal-rich and $\alpha$-poor stars. For this
population, all three velocity components have negative gradients over
[Fe/H]. Among these, the slope of the \vph\ component is the steepest
($\partial V_{\rm \phi}/\partial$\feh\ = --28.2 \kmskpc). The gradient of
the velocity dispersion is very shallow for all three velocity
components. The eccentricity distribution is skewed toward
high-$e$ stars. There is a general tendency of the eccentricity to
decrease with decreasing [Fe/H] and increasing \afe.
Furthermore, there are different kinematic characteristics between the
SMRTN and MPTN stars, as addressed below.\\

(a) SMRTN -- Defined by the thin-disk stars with \feh\ $>$ $+$0.1. This
population exhibits slower rotation, slightly higher $e$, and smaller \rperi\
with respect to the MPTN stars. Half of the SMRTN stars are found with $e <$ 0.2.
Their median \rperi\ value is $\sim$ 7 kpc. 

(b) MPTN -- Consists of thin-disk stars with \feh\ $<$ --0.3. This group
of stars rotates relatively faster, and exhibits high $\sigma_{V_{\rm Z}}$
and \zmax, comparable to the most metal-rich thick-disk stars,
particularly for those in the $\alpha$-rich region. The stars in the 
low-\vph\ velocity group (G1) have shorter \rperi\ than the high-\vph\ 
velocity stars. The stars with \vph\ $>$ 240 \kms\ have relatively 
high $e$ ($>$ 0.2).

\item Thick disk -- Comprises chemically metal-poor and $\alpha$-rich stars. 
All three velocity components have positive gradients with [Fe/H]. Among them, 
the \vph\ component exhibits the largest gradient ($\partial V_{\rm
\phi}/\partial$\feh\ = $+$36.9 \kmskpc). The dispersions of all three
velocity components decline with increasing metallicity. A continuous
rise in \zmax\ with increasing \afe\ is found. The eccentricity distribution
becomes broader as decreasing [Fe/H] or increasing \afe.
	
\end{enumerate}

\subsection{The Super Metal-rich Thin Disk}

In Section \ref{sec:ecc}, we noted that 27\% of our SMRTN stars have low
eccentricity and \rperi\ $>$ 7 kpc. This behavior is confirmed by
other studies. For example, \citet{Kordopatis15} reported that even
though there are high-eccentricity ($e >$ 0.3) stars among their
thin-disk stars, most super metal-rich stars have $e\ \leq$ 0.15 and
large orbital radii. \citet{Hayden18} claimed that the metal-rich stars
with [Fe/H] $>$ $+$0.1 and \rperi\ $>$ 7 kpc account for 25\% of the
stars in their sample, and many stars show low orbital
eccentricity. Considering their large \rperi, these stars 
do not appear to have been formed in a high-metallicity environment 
such as the inner disk, because they never had the chance to move in 
the inner-disk region. However, because these stars have low eccentricity, and 
the metallicity of the interstellar medium (ISM) in the solar neighborhood (SN) is 
roughly [Fe/H] = 0, as \citet{Hayden18} claimed, it is plausible to 
conjecture that these stars in our sample could have been brought from the 
inner disk into the SN by churning. The more recent study 
of \citet{Hayden20} came to the same conclusion, using stars obtained from GALAH. 
The relatively slow rotation velocity of these stars also support migration from
the inner disk by churning. 

Theoretically, \citet{Minchev13} demonstrated that stars born in the
region of 3 $<$ $R_0$ (birth radius) $<$ 5 kpc can contribute to an
increase in the metallicity in the SN by up to 0.6 dex
through radial migration, while stars that are born locally end up with \feh\ $\sim$
0.15, implying that the SMRTN stars were born in the inner disk rather
than the SN.

The above claim is supported by the trend of the velocity dispersion
over [Fe/H] for the SMRTN stars as shown in Figure \ref{fig:7}. The
figure indicates that the gradient of $\sigma_{V_{\rm Z}}$ is slightly
negative, while a small positive gradient for $\sigma_{V_{\rm \phi}}$.
The lower $\sigma_{V_{\rm Z}}$ at the solar metallicity can be explained
by the bias that the stars close to the Galactic mid-plane have more
tendency to migrate because they have relatively lower vertical velocity
dispersion than the ones further away from the Galactic mid-plane.
Consequently, the migrated stars do not show high velocity dispersion,
but rather cooler dispersion \citep{Vera14,Vera16}. The positive
gradient of $\sigma_{V_{\rm \phi}}$ can be explained by the spread of the
changed angular momenta of the migrated star.

How can we interpret the stars with high eccentricity and small
perigalacticon distances (\rperi\ $<$ 5 kpc) among the SMRTN stars?
These stars account for about 22\% of the SMRTN population. The broader
observed $e$-distribution for the low $\alpha$-stars in
Figure~\ref{fig:8} is due to the presence of these stars. As a result of
their shorter perigalacticon distances, these stars are mostly found at
the apogalacticons of their orbits, due to their slower (low-\vph\
velocities) orbital motions there. Therefore, it is plausible to explain that 
the SMRTN stars with high eccentricity came from the inner disk by 
blurring. This interpretation is upheld by studies of \citet{Pompeia02} and
\citet{Trevisan11}, who also claimed that metal-rich stars on
non-circular orbits have come from the inner disk.

\subsection{The Metal-Poor Thin Disk}

Provided that there exists a general increasing trend of stellar age
with increasing \afe, as most recent studies suggest (\citealt{Delgado19, Nissen17, Silva18}), 
we can suppose that the MPTN stars are older than the SMRTN stars 
because of their higher \afe, in spite of their undetermined ages. This group 
of stars in our sample also has mostly high orbital velocity (\vph\ $>$ 180 \kms), 
low-eccentricity ($e <$ 0.3), and large \rperi\ ($>$ 7 kpc), as seen in Figure
\ref{fig:10}. One notable feature in this group of 
stars is the existence of stars with high \vph\ ($>$ 240 \kms) and $e$ $>$ 0.2. 
Furthermore, Figure \ref{fig:12} indicates that the MPTN stars show 
slightly lower \zmax\ and $\sigma_{V_{\rm Z}}$ than those of 
the \afe-poor thick-disk stars.

One possible interpretation on the origin of the high-\vph\ stars among the ones 
with $e > 0.2$ is the radial movement by churning from the outer disk, as these stars have
high-\vph, large \rperi, and lower metallicity than the ISM in the SN. 
In this regard, \citet{Haywood13} proposed that the oldest
thin-disk stars appear to be of outer-disk origin, as they rotate
rapidly with nearly circular orbits. According to their study, there is
a barrier between the inner and outer disks due to outer Lindblad
resonance (\citealt{Lynden-Bell72}), resulting in the isolated and
disconnected evolution of the outer disk from the inner disk. Unless
some of the high-\vph, MPTN stars have moved from the outer disk as
proposed by Haywood et al., we would not observe those stars in the
SN today.

The higher \vz\ dispersion of the MPTN stars compared with the SMRTN
stars also supports the radial migration hypothesis, as the inward
migrators tend to increase the \vz\ dispersion of such stars at their
final destination (\citealt{Vera14}). The negative slope of the \vph\
component over [Fe/H] also provides another clue to the radial
migration, because the stars that arrive from the outer disk should have
higher angular momentum than the ones in the final location of the
migrated stars.

However, the G1 stars ([Fe/H] $<$ --0.3 and low-\vph), required another
mechanism to explain their high eccentricities. Owing to their small
\rperi, they are mostly observed at apogalacticon. Thus, we can
infer from the low metallicity, high-\afe, and relatively low-\vph\ of
these stars that they formed early in the inner disk when the ISM was
not yet enriched with heavy elements, and moved outward due to 
blurring to reach the Solar circle. 

As \citet{Rojas16} suggested, this population might experience
independent chemodynamical evolution, distant from that of their metal-rich
counterparts. They reported that the smooth change of \z\ with [Fe/H]
implies that the dynamical evolution occurred quietly without much 
violent events such as a merger. However, the steep increase of \zmax\ and
$\sigma_{V_{\rm Z}}$ with \afe\ may suggest a different evolution,
as the G1 stars may have been influenced by dynamical heating by a minor
merger. $Gaia$-Enceladus (\citealt{Helmi18}) or the Splashed
Disk (\citealt{Belokurov20}) may be good examples of such mergers. 
  
To summarize, our results suggest that the 
high-metallicity thin-disk stars moved from the inner disk, 
while the low-metallicity thin-disk stars migrated from the outer disk, 
or from the inner disk, through more churning than blurring. 
In line with this interpretation, using red giant stars from APOGEE
DR14 (and $Gaia$ parallaxes), \citet{Feltzing20} attempted to 
quantify the degree of radial migration by blurring and churning. 
They found that about 10--50\% of their sample experienced churning or 
blurring over the metallicity range from [Fe/H]
$\sim$ --0.7 to $\sim$ $+$0.3, even though the fraction varies depending on
the constraints imposed from the circularity of their orbits. 
In addition, a recent study by \citet{Frankel20} also
demonstrated that the evolution of the $\alpha$-poor thin-disk stars is more 
influenced by churning, which agrees with our interpretation. From this line of 
reasoning, we can conclude that radial mixing by churning 
and blurring may have had a more dominant influence in shaping the current 
thin-disk structure of the MW, compared with other suggested mechanisms such as 
accretion and/or disk-heating by mergers.

 \subsection{The Thick Disk}

There is clear kinematic distinction between the thin and thick disks -- the continuous
decrease of $\sigma_{V_{\rm R}}$, $\sigma_{V_{\rm \phi}}$, and $\sigma_{V_{\rm Z}}$ 
with increasing \feh\ for the thick disk, but an almost flat gradient
over metallicity for the thin disk, as shown in Figure \ref{fig:7}. The most notable 
kinematic difference between the two populations is the gradient of the rotation 
velocity over [Fe/H]. We found $\partial V_\phi$/$\partial$\feh\ = --28.2 \kmsdex\ and 
$+$36.9 \kmsdex\ for the thin and thick disk, respectively. The trend of
higher eccentricity with decreasing metallicity is also characteristic
of the thick-disk population. These are tell-tale signs of the different
formation histories between the two disks. Below, we attempt to
associate our findings with the proposed formation models of the thick
disk.

The positive gradient of the rotation velocity is mostly reproduced by
various radial migration models. For instance, \citet{Loebman11} carried
out a simulation of radial migration and identified $\partial
V_\phi$/$\partial$\feh\ $\sim$ --24.8 \kmsdex\ for their low-\afe\
stars, while a mild positive gradient ($\sim$ $+$13.5 \kmsdex) was obtained
for their high-\afe\ stars, although this gradient is substantially
smaller than observed for the thick-disk population. \citet{Curir12}
investigated the evolution of the thick disk by simulated collisionless
particles, and found a positive gradient of $+$60 \kmsdex\ for the thick
disk, which is comparable with the observational result. Recent studies
(\citealt{Kawata17,Schonrich17}) also demonstrated that the inside-out
and upside-down formation of the thick disk, along with radial
migration, can readily reproduce the observed positive slope of orbital
rotation velocity with [Fe/H]. 

Moreover, as shown in Figure \ref{fig:8}, we observe that the eccentricity
distribution of the thick disk varies according to \afe\ and \feh, and
the number of stars with high eccentricity is reduced with decreasing
\afe\ or increasing \feh. This behavior is reproducible by the radial
migration model (\citealt{Sales09}), which produces relatively small
numbers of high-eccentricity stars for the thick-disk component,
consistent with the shape of the $e$-distribution of our sample. These
results point to a radial migration as an important part of the
thick-disk formation history. Nonetheless, \citet{Minchev12} suggested
that the impact of radial migration on the thickening the disk is not
significant.

In Figure \ref{fig:6}, we noted that the velocity dispersions ($\sigma_{V_{\rm R}}$,
$\sigma_{V_{\rm \phi}}$, and $\sigma_{V_{\rm Z}}$) decline continuously
with increasing \feh. In addition, the top panel of Figure \ref{fig:12}
indicates that \zmax\ continuously declines with decreasing \afe. From
these characteristics, it is challenging to imagine that the thick-disk 
stars were accreted, as they would be expected to have similar velocity
dispersions. Rather, we can infer that the star-forming environment
could have stabilized with decreasing \afe\ during the thick-disk formation
phase, if we use \afe\ as a proxy for a stellar age in the thick disk. 
A tight correlation has been reported between \afe\ and age for 
the thick-disk population \citep{Haywood13,Bensby14}. However, caution
may need to be taken, since the recent study by \citet{Silva18} found no
correlation between age and \afe\ for the thick-disk population, based
on a sample of red giants with asteroseismic ages. These authors also
reported that the velocity dispersion is not well-correlated with age
for the high-$\alpha$ stars. One pitfall of this study is that the
uncertainty of their asteroseismic age is rather large. The median 
uncertainty is about 28.5\%, and becomes larger with increasing age. In
their independent study, \citet{Spitoni19} claimed that the overlapping
ages between the thin- and thick-disk stars arose from uncertainties in
the asteroseismic ages. Thus, the existence of any correlations between
age and \afe\ must be confirmed in the future with better-constrained
age determinations.   

Nevertheless, the observed kinematic features can be seen in the gas-rich 
merger models. According to \citet{Brook04,Brook07,Brook12}), the thick disk 
can be formed from accreted gas during a chaotic merger period. This model also 
predicts that cooler kinematics (e.g., low velocity dispersions) appear to 
emerge for the younger disk stars formed by gas-rich mergers at later times; 
$\sigma_{V_{\rm Z}}$ of stars formed at $z$ = 1.0 are smaller than
the velocity dispersions of stars formed at $z$ = 1.5. The epochs at $z$ =
1.0 and $z$ = 1.5 are the dominant phases of thick-disk formation in
the simulation by \citet{Brook12}. These can be interpreted such that
the thick disk becomes stabilized with decreasing age, which
qualitatively agrees with the bottom panel of Figure \ref{fig:12}, if we
consider \afe\ to be a stellar-age proxy. Furthermore, the thick-disk
stars in our sample tend to have various eccentricities, from low to
high, for all metallicity bins, as shown in Figure \ref{fig:11}. In particular,
the diversity of the eccentricity distributions with a wide range of the
\vph\ velocity for the most metal-poor bin (top-left panel of Figure
\ref{fig:11}) probably indicates a hot and violent state when they were
born, or at least the kinematic evolution of those stars. One downside of
this mechanism is, however, the difficulty of reconciling the very
high-$e$ stars produced by the model with the absence of such high-$e$ stars
in the observed data. \citet{Sales09} suggested that the gas-rich merger
scenario can make stars on non-circular orbits that extend to $e\ \sim$
0.8. 

Nonetheless, the cosmological hydrodynamical simulations performed by
\citet{Buck20} support the gas-rich merger model as a thick-disk
formation model, demonstrating that the low- and high-$\alpha$ sequences
observed in the Galactic disk naturally arise through a gas-rich merger.
In the simulations, the high-$\alpha$ stars formed first, and the
low-$\alpha$ stars formed after fresh metal-poor gas was injected into
the disk by the merging satellite. These simulations also showed that
the disk stars were greatly influenced by secular evolution, such as
migration, that is, the low-$\alpha$ stars came from both the inner and
outer disk, whereas the high-$\alpha$ stars mostly migrated from the
inner disk, which qualitatively agrees with our interpretation.

In line with the gas-rich merger models, it is worth mentioning
that a thick disk formed from turbulent clumps can have stars on highly
eccentric orbits, as gravitational instabilities make stars and gas
disperse during the clumpy phase (e.g., \citealt{Bournaud09,Robin14,
Bournaud14,Elmegreen17}). The notable feature of this scenario is the
decreasing scale height of the thick disk over time, which is qualitatively in
good agreement with the observations (see Figure \ref{fig:12}).

There is, however an assertion that the metal-poor and \afe-rich stars
exhibit relatively constant kinematic trend 
(e.g., \citealt{Minchev14}). This constant but high velocity dispersion
can be observed in our sample (bottom panel of Figure \ref{fig:12});
there exists a plateau (or only slightly downward trend) of
$\sigma_{V_{\rm Z}}$ for the \afe\ rich bins. To confirm this further,
we divided our sample into several metallicity bins, and then, checked
the velocity dispersion as a function of \afe\ again. In this exercise,
we obtained a decreasing velocity dispersion in the range of --1.2 $<$
\feh\ $<$ --0.9. This behavior can be seen at the lowest [Fe/H] bin in
Figure \ref{fig:7}. \citet{Minchev14} proposed that the decreasing
velocity dispersion over \afe\ arises from radial migration, triggered
by a merger with a mass ratio of 1 to 5 about 8 -- 9 Gyrs ago. They
emphasized that an external event such as a significant merger is
necessary to trigger the radial migration rather than just the
operation of an internal, secular process.

Recent studies also indicate that the dynamical heating of the disk by
mergers contributes to the formation of the thick disk. For example,
\citet{Belokurov20} used the $Gaia$ DR2 and various spectroscopic data
to identify a large number of metal-rich ([Fe/H] $>$ --0.7) stars on
highly eccentric orbits. They claimed that this population, referred to
as the ``Splashed Disk'', is linked to the $\alpha$-rich thick disk, and has
halo-like kinematics. They argued that the Splashed-Disk stars were born
in the proto-disk of the MW, and their orbits were changed by a massive
ancient accretion event such as $Gaia$-Enceladus. However, their
Splashed-Disk stars have much lower rotation velocities than our
thick-disk population. \citet{An20} confirmed these properties, 
that the Splashed-Disk stars show similar metallicity to the canonical 
thick disk, but their kinematics differ. \citet{Helmi18} also showed, 
from $Gaia$ DR2 and APOGEE DR14 data, that a dwarf satellite whose debris contributes
significantly to the inner-halo population, caused the dynamical heating
of a pre-existing disk to form the thick disk.

To sum up, the kinematic and chemical characteristics observed for our
thick-disk stars suggest that the thick disk is formed by a turbulent
gas-rich merger in an inside-out fashion, as the cosmological
simulations by \citet{Schonrich17} and \citet{Kawata17} demonstrated.
During the mergers and later evolution, the dynamical heating of the
pre-existing disk and stellar radial migration have influenced the
evolution of the thick disk further. This interpretation 
partly agrees with the results from recent cosmological simulations by 
\citet{Grand20}. They showed that the progenitor of \gaia-Enceladus 
and the \gaia-Sausage induced gas-rich mergers to form the 
high $\alpha$-thick disk stars, and heated the proto-Galactic disk 
to generate the positive correlation between \vph\ and [Fe/H]. Consequently, 
it appears that various physical mechanisms have played roles in structuring 
the current thick disk; a single formation scenario or mechanism is not able 
to explain all the observed properties.

\section{Summary}\label{sec:summary}

We have made use of SEGUE G- and K-type dwarfs to 
investigate the chemodynamical properties of the Galactic disk system, 
after separation into an $\alpha$-poor thin and an $\alpha$-rich thick disk. 
The dynamical properties of the thin-disk stars indicate that 
some fraction of the MPTN stars radially move from the outer disk or 
the inner disk, while some fraction of the SMRTN stars may originate from 
the inner disk. Radial migration is required to explain 
these phenomena. Churning, in particular, may have greatly influenced to
create the current structure of the thin disk, while we cannot neglect
the importance of the blurring for the high-eccentricity stars. 

According to our results, the formation of the thick disk can be explained 
by violent mechanisms, such as gas-rich mergers or giant turbulent clumps, 
as these models can readily produce significant numbers of high-eccentricity stars,
as well as the thinner and more stable thick disk with decreasing \afe\ (or increasing \feh). 
During the merger, and the later evolution, the dynamical heating of 
a pre-existing disk, as well as radial migration, may both play roles in 
forming the current thick disk. 


\acknowledgments
We thank an anonymous referee for his/her careful review of this paper
to improve the clarity of the presentation.

Funding for SDSS-III has been provided by the Alfred P. Sloan Foundation, the
Participating Institutions, the National Science Foundation, and the U.S.
Department of Energy Office of Science. The SDSS-III Web site is
http://www.sdss3.org/.

Y.S.L. acknowledges support from the National Research Foundation (NRF) of
Korea grant funded by the Ministry of Science and ICT (No.2017R1A5A1070354
and NRF-2018R1A2B6003961). T.C.B. acknowledges partial support for this work
from the grant PHY 14-30152; Physics Frontier Center/JINA Center for the Evolution
of the Elements (JINA-CEE), awarded by the US National Science
Foundation.

\bibliographystyle{aasjournal}


\end{document}